\documentclass[%
 reprint,
 amsmath,amssymb,
 aps,
]{revtex4-2}

\usepackage{graphicx}% Include figure files
\usepackage{dcolumn}% Align table columns on decimal point
\usepackage{bm}% bold math
\usepackage{todonotes}
\usepackage[version=3]{mhchem}
\usepackage{url}
\usepackage[normalem]{ulem}

\usepackage[percent]{overpic}

\usepackage{graphicx,xcolor}
\graphicspath{ {./plots/} }

\usepackage{bbm}
\usepackage[vlined,ruled]{algorithm2e}

\DeclareMathOperator*{\argmax}{arg\,max}
\DeclareMathOperator*{\argmin}{arg\,min}

\newcommand{\R}{\mathbb R}
\newcommand{\C}{\mathbb C}
\renewcommand{\P}{\mathbb P}
\newcommand{\V}{\mathbb V}
\newcommand{\bmu}{\bm\mu}

\newcommand{\sB}{\mathsf B}
\newcommand{\sH}{\mathsf H}
\newcommand{\cO}{\mathcal O}
\newcommand{\cOrb}{\mathcal O_N}
\newcommand{\sO}{\mathsf O}

\newcommand{\sObrbrp}{\sO_{\br,\brp}}
\newcommand{\sOr}{\mathsf O_r}
\newcommand{\sOrr}{\mathsf O_{r,r'}}
\newcommand{\Nsites}{{N_{\rm sites}}}
\newcommand{\Xitrain}{\Xi_{\rm train}}
\newcommand{\Xitest}{\Xi_{\rm test}}
\newcommand{\myspan}{\mbox{span}}

\newcommand{\calH}{\mathcal H}
\newcommand{\calL}{\mathcal L}
\newcommand{\calN}{\mathcal N}
\newcommand{\calO}{\mathcal O}
\newcommand{\calS}{\mathcal S}

\newcommand{\nf}{n_{\mathsf f}}
\newcommand{\nx}{N_x}
\newcommand{\ny}{N_y}

\newcommand{\gi}{n}
\newcommand{\ngi}{^{(\gi)}}
\newcommand{\Ngi}{^{(N)}}

\newcommand{\sBn}{\mathsf B_\gi}
\newcommand{\sUn}{\mathsf U_\gi}
\newcommand{\sBnp}{\mathsf B_{\gi+1}}

\newcommand{\srb}{_{\sf rb}}
\newcommand{\phirb}{\bm\varphi\ngi\srb}

\newcommand{\phirbso}{\varphi\ngi_{{\sf rb},1}}
\newcommand{\phirbsi}{\varphi\ngi_{{\sf rb},i}}
\newcommand{\phirbsm}{\varphi\ngi_{{\sf rb},m}}
\newcommand{\Phirb}{\bm\Phi\ngi\srb}

\newcommand{\lambdarb}{\lambda\ngi\srb}

\newcommand{\lambdarbN}{\lambda\srb}

\newcommand{\phirbsNi}{\varphi_{{\tt rb},i}}
\newcommand{\phirbN}{\bm\varphi\srb}
\newcommand{\PhirbN}{\bm\Phi\srb}

\renewcommand{\sb}{\mathsf b}
\newcommand{\sHrb}{\mathsf h}
\newcommand{\sHrbq}{\mathsf h_{q}}
\newcommand{\sHrbqqp}{\mathsf h_{qq'}}

\newcommand{\sOrbr}{\mathsf o_r}

\newcommand{\lacumo}{\ce{La4Cu3MoO12}}
\newcommand{\res}{\mathsf{Res}}
\newcommand{\valerror}{{\sf err}_{\sf val}}
\newcommand{\vecerror}{{\sf err}_{\sf vec}}
\newcommand{\sferror}{{\sf err}_{\sf sf}}

\newcommand{\bk}{{\bf k}}
\newcommand{\br}{{\bf r}}
\newcommand{\brp}{{\bf r'}}
\newcommand{\bx}{{\bf x}}
\newcommand{\by}{{\bf y}}
\newcommand{\bS}{{\bf S}}

\begin{document}
%%%%% title : short title may not be used but TITLE is required.
\title{Surrogate models for quantum spin systems based on reduced order modeling }

%%%%%%%%
\author{Michael F. Herbst}
\email{herbst@acom.rwth-aachen.de}
\affiliation{%
 Department of Mathematics, RWTH Aachen Unviersity, Schinkelstr. 2, 52062 Aachen, Germany 
}
%%%%%%%%
\author{Stefan Wessel}
\email{wessel@physik.rwth-aachen.de}
\affiliation{%
Institute for Theoretical Solid State Physis, RWTH Aachen University, Otto-Blumenthal-Str.~26, 52074 Aachen, Germany 
}
%%%%%%%%
\author{Matteo Rizzi}%
\email{m.rizzi@fz-juelich.de}
\affiliation{%
Forschungszentrum J\"ulich GmbH, Institute of Quantum Control,
Peter Gr\"unberg Institut (PGI-8), 52425 J\"ulich, Germany;
}
\affiliation{%
Institute for Theoretical Physics, University of Cologne, D-50937 K\"oln, Germany
}%
%%%%%%%%
\author{Benjamin Stamm}
\email{best@acom.rwth-aachen.de}
\affiliation{%
 Department of Mathematics, RWTH Aachen Unviersity, Schinkelstr. 2, 52062 Aachen, Germany 
}
%%%%%%%%

%%%%% Begin Abstract %%%%%%%%%%%
\begin{abstract}
We present a methodology to investigate phase-diagrams of quantum models based
on the principle of the reduced basis method (RBM).
The RBM is built from a few ground-state snapshots, i.e., lowest eigenvectors
of the full system Hamiltonian computed at well-chosen points in the parameter
space of interest. We put forward a greedy-strategy to assemble such
small-dimensional basis, i.e., to select where to spend the numerical effort
needed for the snapshots.
Once the RBM is assembled, physical observables required for mapping out the phase-diagram (e.g., structure factors) can be computed for any parameter value with a modest computational complexity, considerably lower than the one associated to the underlying Hilbert space dimension.
We benchmark the method in two test cases, a chain of excited Rydberg atoms and a geometrically frustrated antiferromagnetic two-dimensional lattice model, and illustrate the accuracy of the approach. 
In particular, we find that the ground-state manifold can be approximated to
sufficient accuracy with a moderate number of basis functions, which increases
very mildly when the number of microscopic constituents grows --- in stark
contrast to the exponential growth of the Hilbert space needed to describe each
of the few snapshots.
A combination of the presented RBM approach with other numerical techniques circumventing even the latter big cost, e.g., Tensor Network methods, is a tantalising outlook of this work.
\end{abstract}
%%%%% end %%%%%%%%%%%

%%%%% AMS/PACs/Keywords %%%%%%%%%%%
%\pac{}
% \ams{XXX}

%%%% maketitle %%%%%
\maketitle

%%%% Start %%%%%%
%%%%%%%%%%%%%%%%%%%%%%%%%%%%%%%%%%%%%%%%%%%%%%%%%%%%%%%%%%%%%%%

% \MFH{British versus American spelling}

\section{Introduction}
\label{sec:Intro}
The study of quantum spin models, dating back to the early days of quantum mechanics, is a central topic in modern condensed matter physics. 
Indeed, as basic quantum many-body systems with  inherently strong correlations, these models often display interesting ground states including complex ordering patterns, quantum disordered regimes or topological order such as in quantum spin liquids~\cite{Broholm2020}. 
In addition, these ground states are in many cases good approximations to the low-temperature behavior of real physical systems or compounds that are described by these models. 
While exact analytical solutions for specific quantum spin models exist, such as the early Bethe-ansatz solution of the one-dimensional Heisenberg spin-$\frac12$ chain~\cite{Bethe1931}, most realistic models instead require advanced computational techniques for their solution.

Most traditionally, a (more or less theory-guided) scan of numerical instances needs to be computed across the parameter space of the Hamiltonian, in order to map out the phase diagram of the model --- commonly through the computation of relevant observables (e.g., structure factors).
Whatever is the method of choice, each such instance is typically very expensive, despite remarkable progresses to get around the naive exponential growth of the Hilbert space with the number of spins in a finite-size sample (see below).
Moreover, the scan could be certainly performed in an embarrassingly parallel manner
and/or initial guesses could be recycled from already converged simulations
for nearby parameter values.
Still, these approaches are burning a considerable amount of CPU hours.
Moreover employing previous simulations as a guess is not unproblematic and may, for example,
give rise to spurious hysteresis in the proximity of phase transitions.

In this paper, we put forward an alternative and complementary strategy, relying on the so-called reduced basis method (RBM) --- borrowed from the numerical mathematics community (see below) --- that promises to tear down the overall amount of expensive computational instances for a faithful phase diagram.
The core idea is to establish cheap surrogate models for a many-query context of parametrized quantum spin models, based on a few sample points for which snapshots, i.e., solutions of the true Hamiltonian problem, are actually produced.
First, in an offline / training phase, these sample points are chosen
--- typically via a greedy strategy --- and a reduced basis is assembled out
of the corresponding snapshots. The aim of the procedure is to obtain a subspace,
which encompasses an accurate approximation of the solution across the given parameter domain.
Then, in a so-called online phase, the special (affine) form of the
parametrized Hamiltonians allows to obtain an approximate solution
for any parameter value in a complexity solely depending on the assembled
low-dimensional space and not on the potentially very high dimension of the
Hilbert space.

Using such small dimensional effective models in engineering, physics and
numerical modeling is an old idea, which has fruited not only in theoretical considerations,
but also computational implementation.
Early contributions in the field of such reduced order modeling~\cite{fox71,noor80,noor82,sirovich1987turbulence} include applications to partial differential equations (PDEs) that originate in structural and fluid mechanics. 
Between 2000 and 2010 the method has gained much popularity due to the mathematically rigorous error control through the variational framework~\cite{prud2002reliable}
alongside which the notion of a reduced basis method (RBM) has been established.
Nowadays, reviews~\cite{rozza2008reduced} as well as monographs~\cite{hesthaven2016certified,quarteroni2015reduced} are available on the topic.
In the context of eigenvalue problems
the application of the RBM is less popular, despite the fact that the idea was
already used in early contributions.
See for example~\cite{aktas1998reduced} for an
application in structural mechanics using an effective small dimensional
basis and following a variational ansatz.
For parametric eigenvalue problems in a PDE setting this idea has then been formalized
in~\cite{machiels2000output} and extended in~\cite{fumagalli2016reduced} using
\textit{a posteriori} error estimators.
In~\cite{horger2017simultaneous} a generalization to
target clusters of eigenvalues of a parametrized eigenvalue problem has been presented.

Here, as anticipated, we want to bring the benefits of the reduced basis method to the
quantum world and exploit it to speed up and economize the parameter scans required
for the generation of quantum phase diagrams.
As a proof of concept, we test the methodology on two quantum-spin models,
namely a model used for a chain of excited Rydberg atoms and a model for the  antiferromagnetically coupled  spin-$\tfrac12$ triangles in the compound \lacumo.
Noticeably, for a given precision target, the number of required snapshots stays
moderate even when the investigated region spans different phases of the
model. Furthermore we will demonstrate this number to grow only weakly
with the system size.
Both features are somehow pleasantly surprising and they open up a number of
theoretical questions. 
Moreover, the RBM strategy could contribute to green computing as well.

Before delving into the technical presentation, let us note that the RBM framework
is agnostic to the precise numerical technique employed for obtaining the snapshots.
It is thus fully complementary to existing algorithms.
To keep this benchmark study simple we will employ exact numerical diagonalization (ED)%
~\cite{Laeuchli2011, Sandvik2010},
which is however strongly limited by the exponential growth of the Hilbert
space with the number of spins in a finite-size sample.
One possibility to evade such restriction is stochastic sampling of the
wavefunction via quantum Monte Carlo (QMC) methods~\cite{Sandvik2010}: however,
in the presence of geometric frustration, quantum Monte Carlo typically suffers
severely from the negative-sign problem~\cite{Troyer2005}.
A modern alternative --- less prone, if not immune, to such issues --- is offered by the density matrix renormalization group (DMRG)~\cite{Schollwoeck2011} and its descendant tensor-network (TN) approaches~\cite{Verstraete2008,Eisert2013,Orus2014,Silvi2019}. 
These are based on the insight that physically relevant states are typically low-entangled and thus occupy only a comparably small subset of the total Hilbert space.
Of note this is a different concept of space reduction compared to the RBM approach
with the relation between the two still remaining to be explored.

The paper is organized as follows. In Section~\ref{sec:Problem}, we formalize the quantum spin problems as abstract parametrized eigenvalue problems and show how the two test examples, i.e., the chain of excited Rydberg atoms and the geometrically frustrated antiferromagnetic two-dimensional lattice model, can be cast in this framework.
Section~\ref{sec:Method} introduces the reduced basis method for the abstract family of model problems with particular focus on handling degenerate states that might appear. Section~\ref{sec:Results} presents the numerical results for the two test cases while Section~\ref{sec:Conclusion} is left for conclusions.

%%%%%%%%%%%%%%%%%%%%%%%%%%%%%%%%%%%%%%%%%%%%%%%%%%%%%%%%%%%%%%%
%%%%%%%%%%%%%%%%%%%%%%%%%%%%%%%%%%%%%%%%%%%%%%%%%%%%%%%%%%%%%%%
\section{Problem setting}
\label{sec:Problem}
We first introduce an abstract form to describe quantum spin models, and later show how our two concrete examples can be cast into this form.
We consider a system with $\Nsites$ degrees of freedom (e.g., quantum spins), such that the quantum state of the total system belongs to the
Hilbert space $\calH = \C^{\calN}$, with $\calN=d^\Nsites$, and where $d$ denotes the dimension 
of the Hilbert space of each of the $\Nsites$ 
degrees of freedom. 
The interactions are modeled by an Hamiltonian $\sH(\bmu)$ formulated in a so-called affine decomposition, i.e.:
\begin{equation}
    \label{eq:affine_decomp}
    \sH(\bmu) = \sum_{q=1}^{Q} \theta_q(\bmu) \, \sH_q \, ,
\end{equation}
where $\bmu\in\P$ denotes a set of parameters in the parameter domain of interest, 
$\theta_q:\P \to \R$ indicates a scalar parameter-dependent function,
and $\sH_q: \calH \to \calH$ are Hermitian matrices of dimension $\calN \times \calN$.
We assume that the number of terms $Q$ is independent of $\mathcal N$.

We are now interested to evaluate, for each $\bmu\in\P$, the ground-state(s) of the Hamiltonian $\sH(\bmu)$, i.e., the eigenvector(s)  $\bm\Psi(\bmu)=(\Psi_1(\bmu),\ldots,\Psi_m(\bmu))$ corresponding to the smallest eigenvalue $\lambda(\bmu)$ solution to the problem 
\begin{equation}
    \label{eq:Hdim_evp}
    \sH(\bmu) \bm\Psi(\bmu) = \lambda(\bmu) \bm\Psi(\bmu) \, .
\end{equation}
Note that our numerical RBM method will naturally allow for ($m$-fold) degenerate ground-states, which can arise due to symmetries.

In most applications, however, one is not so much interested in the high-dimensional solution $\bm\Psi(\bmu)$ itself, but rather in the (scalar) expectation values of a set of physical observables as the parameters are tuned around, i.e., into a collection $\calO(p):\P \to \R$ of functionals of $\lambda(\bmu)$ and $\bm\Psi(\bmu)$ ($p$ being the index running over the collection).
Although a complete abstract framework is possible, we restrict ourselves to affine-decomposable observables, i.e., 
\begin{equation}
    \label{eq:affine_decomp_output1}
    \sO(p) = \sum_{r=1}^{R} \alpha_r(\bmu;p)\, \sOr \, ,
\end{equation}	
where \mbox{$\alpha_r(\bmu;p)\in\C$} are scalar coefficients and $\sOr$ parameter-independent matrices of dimension $\calN\times\calN$.
Furthermore, we consider only averages over the ground-manifold:
\begin{equation}
    \label{eq:affine_decomp_output2}
    \calO(\bmu;p) = \sum_{r=1}^{R} \alpha_r(\bmu;p) \frac{1}{m} \sum_{i=1}^m\Psi_i(\bmu)^\dagger \sOr \Psi_i(\bmu) \, .
\end{equation}
In the applications below, for example, $p$ will take the role of the Fourier wave-vector for the structure factor as observable.

We will address the particular scenario where the solution to these eigenvalue problems need to be solved for many different parameter values $\bmu$ in a many-query context.
For example a sweep through the parameter space, i.e., evaluating the map $\bmu \mapsto \cO(\bmu;p)$ for many parameter values $\bmu\in\P$.
Despite the fact that state-of-the-art numerical methods do not scale exponentially with the number of constituents
--- in contrast to the Hilbert-space dimension ---
a fine-resolved scan over the parameter domain is still very expensive
due to the overall complexity to compute the ground-state \emph{for each new parameter} value.
In such cases of a many-query context, we will make here use of the linear dependency of the different solutions (from one parameter value to the other)
and thus render such computations considerably more affordable.

Let us now pass to explicitly show how two physical problems of current interest in the community can be cast in this abstract framework.

%%%%%%%%%%%%%%%%%%%%%%%%%%%%%%%%%%%%%%%%%%%%%%%%%%%%%%%%%%%%%%%

\subsection{Chains of excited Rydberg-atoms}
\label{ssec:rydberg}
As a first example, we consider Rydberg atoms assembled into a regular lattice by means of optical tweezers, as it became customary in recent years, in a series of experiments of increasing relevance for quantum simulation purposes~\cite{schauss2012observation,labuhn2016tunable,bernien2017probing,Schauss2018Ising,morgado2021}.
We pick up here the simplest  -- yet very insightful --- setup of an equally-spaced one-dimensional chain (with lattice spacing $a$), with atoms modeled as two-level systems coupled by an external laser with Rabi frequency $\Omega$.
The interplay between the level-detuning $\Delta$ and the dipolar interaction strength between excited atoms gives rise to a wealth of different breaking patterns of the (discrete) translational symmetry, dictated by the Rydberg-blockade radius $R_\mathrm{b}$~\cite{Fendley2004CDW,Pohl2010Crystal,rader2019floating,chepiga2021lifshitz,chepiga2021kibble,keesling2019kibble}.

We therefore consider the following Hamiltonian, scaled by $\hbar \Omega$ (and with $n_S=R_b/a$):
\begin{equation} \label{eq:HamRydberg}
    \mathsf H(\bmu) :=
    \frac{1}{2} \sum_r \sigma_r^x -\frac{\Delta}{\Omega} \sum_r \hat n_r + \sum_{r<r'} \left(\frac{n_S}{r'-r}\right)^6 \hat n_r\hat n_{r'} \, ,
\end{equation}
where the indices $r,r'$ belong to the range $\calL = \{1,\ldots,\nx\}$. Here, $\sigma^x_r$ and $\hat n_r=\frac12 (1+\sigma^z_r)$ denote, respectively, the $x$-Pauli matrix and the Rydberg excitation (single particle) operators corresponding to the site $r$, according to the common convention:
\begin{equation} \label{eq:tensop}
     A_r := \left(\otimes^{r-1} \mathbbm{1}\right) \otimes A \otimes  \left(\otimes^{\nx-r} \mathbbm{1}\right) \, ,
\end{equation}
for  any operator $A$ acting on a single site.

Evidently, the Hamiltonian~\eqref{eq:HamRydberg} is already formulated as an affine decomposition, see Eq.~\eqref{eq:affine_decomp}, with  
\begin{equation*}
    \theta_1(\bmu) = 1
    \quad
    \theta_2(\bmu) = -\mu_1 = -\frac{\Delta}{\Omega}
    \quad
    \theta_2(\bmu) = \mu_2^6 = n_S^6 \, ,
\end{equation*}
the corresponding Hamiltonian components $\sH_{q \in \{1,2,3\}}$ and the parameter vector $\bmu=(\Delta/\Omega,n_S)$ being easily identified.

The identification of the different symmetry-breaking patterns is particularly transparent by looking at the structure factor,
\begin{equation}\label{eq:StructFactRy}
  \calS(\bmu;k) 
    = \frac{1}{\nx} \sum_{r,r'}  \exp\Big(- \mathrm{i} \,(r-r') k \Big) \,\langle \hat n_r \hat n_{r'} \rangle_{\bmu} \, ,
\end{equation}
which therefore will constitute the (parametrized) output functional we are interested in.
Let us notice that this is also already in the affine decomposition form assumed in Eq.~\eqref{eq:affine_decomp_output2}, by introducing
\begin{equation}\label{eq:StructFactRy_param}
    \alpha_{r,r'}(k) = \frac{\exp\Big(- \mathrm{i}\,(r-r')k\Big)}{\nx},
    \quad \mbox{and } 
    \sOrr = \hat n_r \hat n_{r'}.
\end{equation}

%%%%%%%%%%%%%%%%%%%%%%%%%%%%%%%%%%%%%%%%%%%%%%%%%%%%%%%%%%%%%%%

\subsection{Antiferromagnetic spin-$\tfrac12$ triangles in \lacumo}
\label{ssec:triangle}

As a second example we consider a quantum \mbox{spin-$\frac12$} system that is composed out of a square lattice of triangular units (trimers), each containing three spins. 
This model was examined previously~\cite{WesselHaas2001,Wang2001} as a basic model to describe the magnetism observed in \lacumo~\cite{Qiu2005,Azuma200}.
Due to an antiferromagnetic coupling of the trimer-spins, this system is geometrically frustrated. As a result, it cannot be efficiently studied by QMC methods, for example, due to a severe sign problem.
In earlier studies~\cite{WesselHaas2001,Wang2001}, its ground-state phase diagram, as a function of varying the intra-dimer couplings (as specified below), has been obtained using exact numerical diagonalization calculations, with one expensive numerical instance per point in the parameter space $\P$.
Below, we will re-examine this model within our reduced order modeling approach, which will allow us to obtain the same ground-state phase diagram through a much cheaper scan over~$\P$ with the surrogate model obtained in a (short) training phase.

\begin{figure}
    \centering
    \includegraphics[trim = 0mm 30mm 0mm 30mm, clip,width=0.48\textwidth]{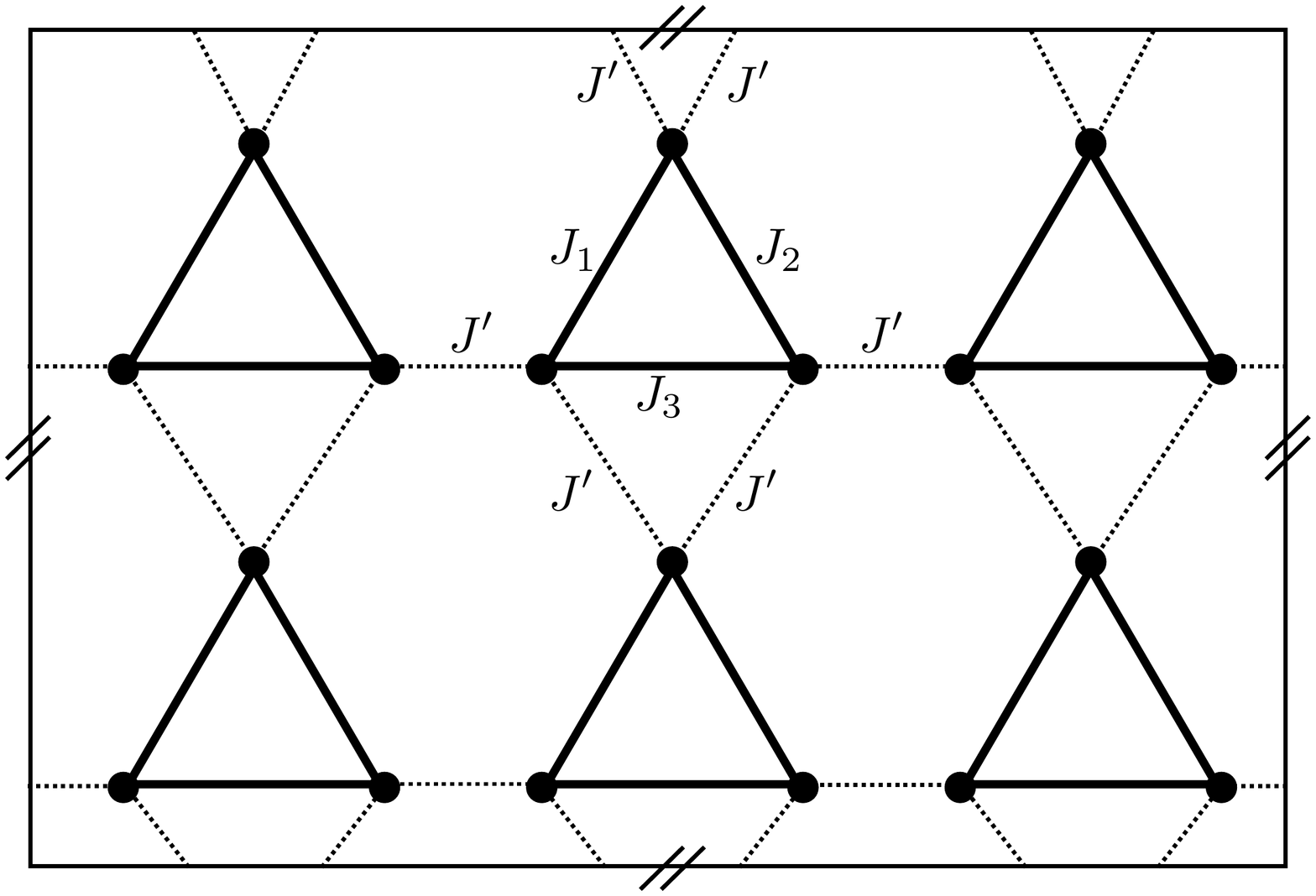}
    \caption{Schematic illustration of the antiferromagnetic spin-$\tfrac12$ triangle model in the case $\nx=3$, $\ny=2$.}
    \label{fig:triangle-model}
\end{figure}

We thus consider the following (scaled) Hamiltonian:
\begin{eqnarray}\label{eq:AFTriangles}
    \sH(\bmu) 
    &= & 
    \sum_{\br} \Big[ 
    \frac{J_{1}}{J_3} \bS_{\br,1}\cdot \bS_{\br,2}
    + \frac{J_{2}}{J_3} \bS_{\br,2}\cdot \bS_{\br,3}
    +  \bS_{\br,3}\cdot \bS_{\br,1}
    \\ \nonumber
    & &
    + \frac{J'}{J_3} \left(
    \bS_{\br,3}\cdot \bS_{\br+\bx,1}
    + \bS_{\br,2}\cdot \bS_{\br+\by,1}
    + \bS_{\br,2}\cdot \bS_{\br+\by,3}
    \right)
    \Big] \, ,
\end{eqnarray}
for $\bmu=(\frac{J_1}{J_3},\frac{J_2}{J_3},\frac{J'}{J_3})$,
where we decided to adopt $J_3$ as the unit of energy.
It is evident that it is already cast in its affine decomposition form of Eq.~\eqref{eq:affine_decomp} with % $Q=4$ and 
\begin{align*}
    \theta_1(\bmu) &= \mu_1 = \frac{J_{1}}{J_3}
    \quad
    \theta_2(\bmu) = \mu_2 = \frac{J_{2}}{J_3}
    \\
    \theta_3(\bmu) &= 1
    \quad
    \theta_4(\bmu) = \mu_3 = \frac{J'}{J_3} \, ,
\end{align*}
and the $Q=4$ Hamiltonian components directly readable from the above.
The position vectors $\br$ belong to a regular $(\nx\times \ny)$-lattice $\calL=\mathbb Z^2\cap [0,\nx)\times [0,\ny)$ with unit cell spanned by the unit vectors $\bx=(1,0)^\top$ and $\by=(0,1)^\top$, basis elements denoted by $\alpha\in \{1,2,3\}$, and periodic boundary conditions imposed as $\br + \nx \bx = \br$ and $\br + \ny \by = \br$ $\forall \br \in \calL$, respectively. 
Finally, $\bS_{\br,\alpha}$ denotes the vector of spin operators $\bS = (S^x, S^y, S^z) = \tfrac{1}{2} (\sigma^x, \sigma^y, \sigma^z)$ acting accordingly to Eq.~\eqref{eq:tensop} on the site $r = (\br,\alpha)$, where we converted tuples into a linear index ranging up to $3\nx \ny$, and
\[
    \bS_{\br,\alpha} \cdot \bS_{\br',\alpha'}
    =
    S_{\br,\alpha}^x S_{\br',\alpha'}^x
    +
    S_{\br,\alpha}^y S_{\br',\alpha'}^y
    +
    S_{\br,\alpha}^z S_{\br',\alpha'}^z.
\]
Figure~\ref{fig:triangle-model} provides a schematic illustration of the model and we refer to~\cite{WesselHaas2001} for more details about the formulation.

The output functional we are interested in is again the structure factor
\begin{equation}\label{eq:StructFactTri}
    \calS(\bmu;\bk) 
    = 
    \frac{1}{\nx \ny}\sum_{\br,\br'} \exp\Big(-\mathrm{i}\, (\br-\brp) \cdot \bk \Big) \langle \overline\bS_{\br}\cdot \overline \bS_{\brp} \rangle_{\bmu},
\end{equation}
with the trimer total-spin $\overline \bS_{\br}  = \overline \bS_{\br,1} + \overline \bS_{\br,2} + \overline \bS_{\br,3}$, see~\cite{WesselHaas2001}.
By introducing 
\begin{equation}\label{eq:StructFactTri_param}
    \alpha_{\br,\brp}(\bk) =
    \frac{\exp\Big(- \mathrm{i}\,(\br-\brp) \cdot \bk\Big)}{\nx \ny},
    \  \mbox{and } 
    \sObrbrp = \overline\bS_{\br}\cdot \overline \bS_{\brp},
\end{equation}
we recover the affine decomposition assumed in~\eqref{eq:affine_decomp_output2}.

%%%%%%%%%%%%%%%%%%%%%%%%%%%%%%%%%%%%%%%%%%%%%%%%%%%%%%%%%%%%%%%
%%%%%%%%%%%%%%%%%%%%%%%%%%%%%%%%%%%%%%%%%%%%%%%%%%%%%%%%%%%%%%%

\section{Method}
\label{sec:Method}
\begin{figure*}[t!]
    \centering
    \begin{overpic}[width=0.45\textwidth]{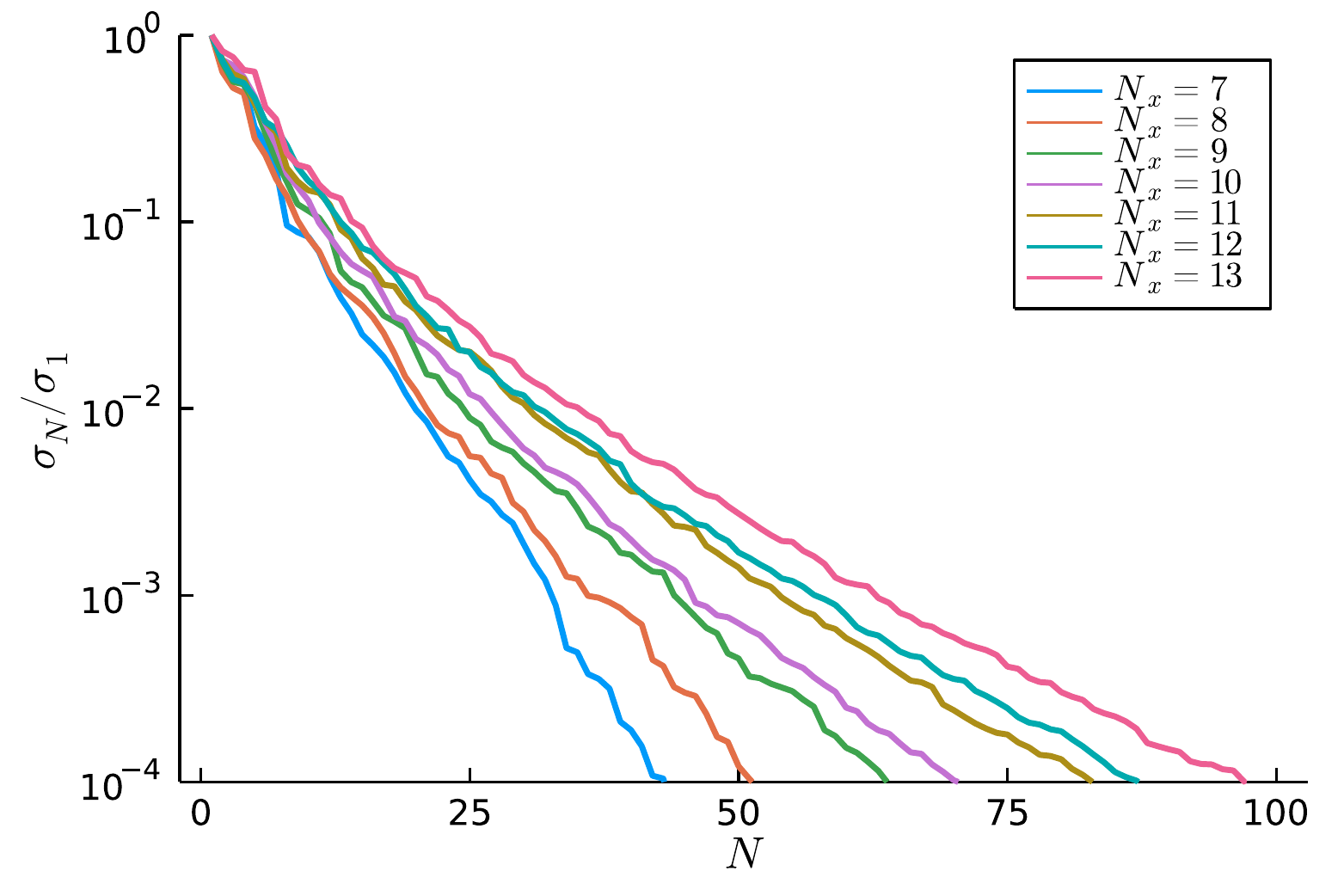}
    \put (0,62) {(a)}
    \end{overpic}
    \begin{overpic}[width=0.45\textwidth]{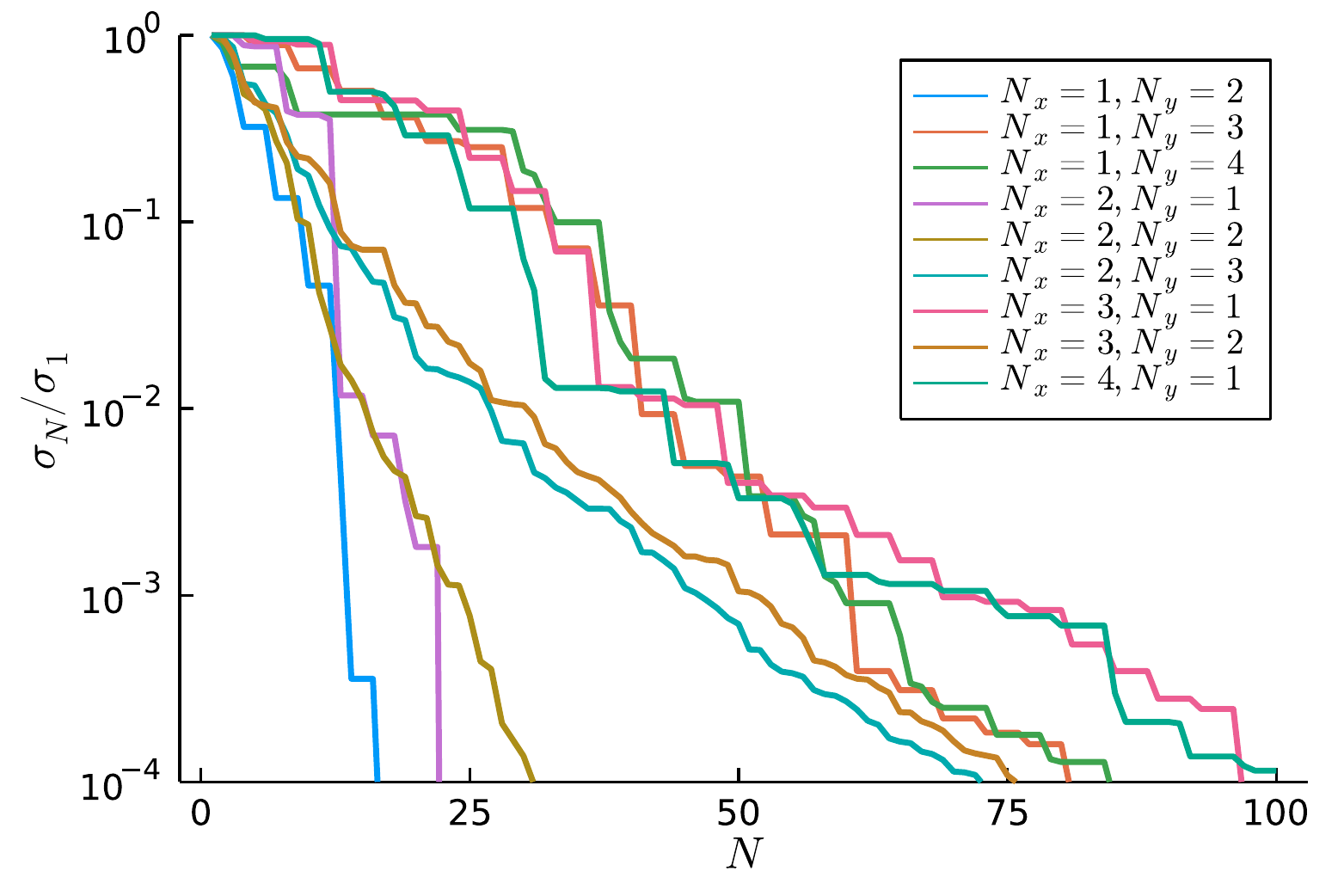}
    \put (0,62) {(b)}
    \end{overpic}
   \caption{
       Decay of normalized singular values of a snapshot matrix $A = \big[\bm\Psi(\bmu_1)\;|\;\cdots\;|\;\bm\Psi(\bmu_{N_M})\big]$ consisting of wave-functions for a given $\Xitest$ consisting of $49\times 49$ uniformly spaced points in $\P=[0, 5]\times[0.5, 4]$ and $9\times 9\times 9$ uniformly spaced points in $\P=[0, 2]\times[0, 2]\times [0.01, 0.1]$ for the case of Rydberg chains (a) and antiferromagnetic spin-$\frac{1}{2}$ triangles (b), respectively.
    }
    \label{fig:svd}
\end{figure*}

\begin{figure}
    \centering
    \includegraphics[width=0.48\textwidth]{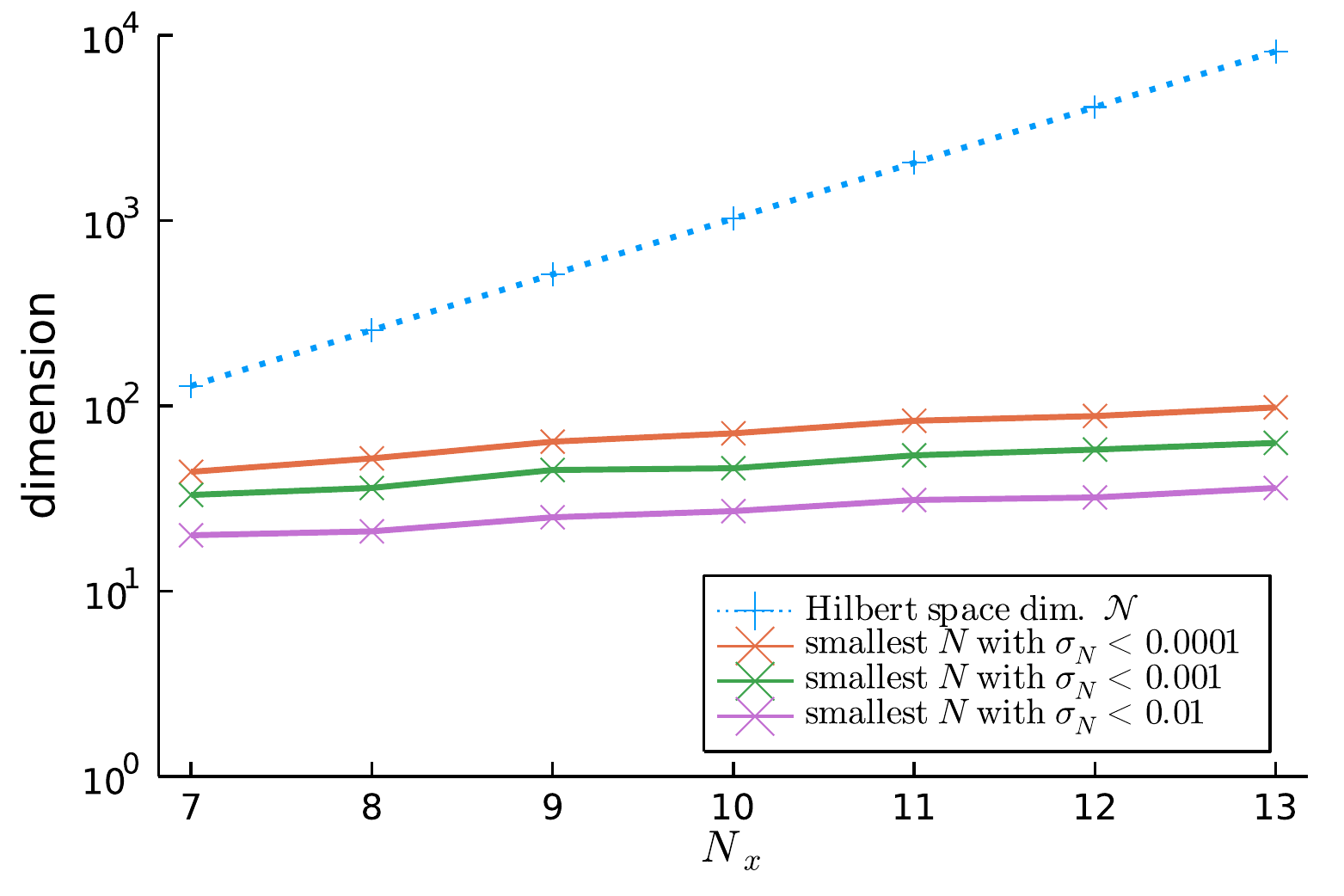}
    \caption{Comparison of basis size to reach particular truncation error
        and size of the Hilbert space.}
    \label{fig:singvalthresh}
\end{figure}

In this work, we want to make explicit use of the fact that the solution may
--- and for our considered examples indeed does
--- exhibit high linear dependency for similar parameter values.
This is a working assumption that shall be verified using the upcoming greedy algorithm for each system individually. 
In such cases,
a low-dimensional basis $\sB$ should exist to describe the overall solution manifold
\[
    \mathcal{M} = \Big\{ \bm\Psi(\bm\mu)\in \calH \;\Big|\; 
    \Psi(\bm\mu) 
    \mbox{ g.s.~of } \sH(\bm\mu),\; \forall
    \bmu \in \P
    \Big\}\subset \calH 
\]
of the (possibly degenerate) ground-states (g.s.) $\bm\Psi(\bmu)$ of the Hamiltonian $\sH(\bm\mu)$ under variation of the parameters $\bm\mu\in\P$, i.e.,
\[
	\myspan \, \sB \simeq \mathcal{M}\, , \mbox{with } \dim \sB \ll \calN.
\]
We postulate that this can be the case even on parameter domains including non-trivial phase transitions.
This working assumption is supported by numerical evidence in Figure~\ref{fig:svd}
for the two problems that we consider here as a benchmark.
We provide the decay of the singular values $\sigma_N$ of a snapshot matrix $A = \big[\bm\Psi(\bmu_1)\;|\;\cdots\;|\;\bm\Psi(\bmu_{N_M})\big]$ over a test-grid $\Xitest$ representing the best approximation error (in the $\ell^2$-sense over $\Xitest$) using $N$ basis functions. The details of the computations are provided in the figure caption.
In the case of the Rydberg chain, we observe that the number of basis functions required to approximate any element of the snapshot matrix for fixed tolerance only increases very mildly for increasing $\nx$.
Note that, relative to the dimension $2^{\nx}$ of the underlying Hilbert space, the solution manifold is very low-dimensional. 
This is illustrated in Figure~\ref{fig:singvalthresh} where the value of $N$, for different tolerances, are plotted with respect to the size of the Rydberg chain with a direct comparison to the dimension $\mathcal N=2^{\nx}$ of the Hilbert space (note the $\log$-scale for the y-axis).
For the frustrated triangular model, instead, we observe an apparent significant increase with respect of the parameters $\nx,\ny$. This can be explained by the fact that, in a very specific region of the parameter domain we are considering --- namely for  $J_1=J_2$ and $J' \ll J_3$ --- we observe degenerate ground-states with degeneracies of degree $2^{\nx\ny}$, and this significantly increases the dimension of the solution manifold targeted by the RBM.
Such behavior is however related to the model- (and actually parameter-) dependent high frustration, and would affect any other approach with a similar impact. We included this parameter regime as a stress-test for our RBM, while we could have circumvented it by considering only larger values of $J'/J$ or a training grid not including the diagonal in the $J_1$-$J_2$ plane.
Of note, even $2^{\nx\ny}$ is asymptotically fairly small
compared to the dimension $2^{3\nx\ny}$ of the Hilbert space,
motivating to pursue a reduced-basis approach even in this setting.

Let us be once more explicit in stating that here we deal with the complexity
in the parameter space, not with the exponential cost in the physical system size.
The latter can be accounted for by combining our method with a numerical method of choice in a black-box manner, whenever the numerical instance for a particular parameter-value needs to be computed.
Our RBM-framework is therefore complementary (and not competitive)
with respect to established numerical many-body methods.

Based on the postulate above, one can develop an efficient method providing an effective surrogate model for cheap scans over the entire parameter domain within a many-query context.
Indeed, once such a low-dimensional approximation is constructed, the ground-state computation can simply be performed by a classical Rayleigh-Ritz procedure in this subspace.
The method described here consist of an offline-online procedure as is classical in the context of the reduced basis method~\cite{hesthaven2016certified}:

\paragraph{Offline:}
This step consists of constructing the low-dimensional approximation space by using a greedy algorithm~\cite{hesthaven2016certified}, which we recall here for the sake of completeness.
Given a so-called training (or trial) grid $\Xitrain$ consisting of
a sizable number of $M$ training points in $\P$, we will generate a sequence of
low-dimensional approximation spaces of the form
\[
    \V_\gi := \myspan
    \Big\{ \bm\Psi(\bmu_{1}),\ldots,\bm\Psi(\bmu_{\gi})\Big\}
    \equiv \myspan \, \sBn\, ,
\]
where each $\bm\Psi(\bmu_{k})=\big(\Psi_1(\bmu_{k}),\ldots,\Psi_{m_k}(\bmu_{k})\big)$, \mbox{$k=1,\ldots,\gi$}, denotes all $m_k$ ground-states of $\sH(\bmu_{k})$,
explicitly including the treatment of degenerate ground-states in the formalism.
The dimension of $\V_\gi$ is therefore $d_\gi:=\mbox{dim}(\V_\gi) \le \sum_{j=1}^\gi m_j$, and $\sBn\in\C^{\mathcal N \times d_\gi}$ denotes a basis,
for which technical details will be given later.
The parameter values $\{\bmu_{1},\ldots,\bmu_{\gi}\}\subset\Xitrain$ (sample points)
are a few well-chosen points,
whose selection criterion will be described shortly as well.
We call $\V_\gi$ the reduced basis space.
The construction of $\V_\gi$ requires the computation of $\gi$ (truth) high-dimensional problems of the form of Eq.~\eqref{eq:Hdim_evp}.

Starting with an initial parameter value $\bmu_{1}\in\Xitrain$, the selection of $\bmu_k$ is performed by induction and is based on a greedy-algorithm.
Thus, let us assume that $\V_\gi$ based on $\{\bmu_{1},\ldots,\bmu_{\gi}\}$ is given.
To select  $\bmu_{\gi+1}$ we proceed as follows. We solve
\begin{equation}
    \label{eq:RB_rayleigh}
    \Phirb(\bmu) = \argmin_{\Phi \in \V_\gi} \frac{\Phi^\dagger \sH(\bmu) \Phi }{\Phi^\dagger \Phi}, 
\end{equation}
for each $\bmu\in\Xitrain$, where $\Phirb(\bmu)$ denotes the vector of (possibly degenerate) ground-states.
Using the variational principle, the above solution can be realized by solving the $d_\gi$-dimensional (generalized) eigenvalue problem
\begin{equation}
    \label{eq:Reduced_evp}
    \sHrb(\bmu) \, \phirb(\bmu) = \lambdarb(\bmu) \, \sb \, \phirb(\bmu)
\end{equation}
with $\sb=\sBn^\dagger \sBn\in\C^{d_\gi \times d_\gi}$.
Here $\lambdarb(\bmu),\phirb(\bmu)$ denote explicitly the smallest eigenvalue
of multiplicity $m$ with corresponding eigenvectors
$\phirb(\bmu)=\big(\phirbso(\bmu),\ldots,\phirbsm(\bmu)\big)$ of $\sHrb(\bmu)$.
The reduced (or compressed) Hamiltonian $\sHrb(\bmu)$ is assembled as
\begin{equation} 
    \label{eq:Hn_assembly}
    \sHrb(\bmu) = \sum_{q=1}^{Q} \theta_q(\bmu) \, \sHrbq,
\end{equation}
with $\sHrbq= \sBn^\dagger \sH_q \sBn\in\C^{d_\gi \times d_\gi}$.
The solutions to Eq.~\eqref{eq:RB_rayleigh} and Eq.~\eqref{eq:Reduced_evp}
are related by the expression $\Phirb(\bmu) = \sBn \phirb(\bmu)$.
In other words $\Phirb(\bmu)$
is the solution represented in the Hilbert space $\calH$
and $\phirb(\bmu)\in\C^{d_\gi \times m}$ collects the coefficients
of all $\Phirb(\bmu)$ when expressed in the basis $\sBn$.
In accordance with the usual terminology of numerical linear algebra
we will also use the terms Ritz vector and Ritz value
to refer to the RBM-approximations $\lambdarb(\bmu)$ and $\Phirb(\bmu)$
of the true ground states $\lambda(\bmu)$ and $\bm\Psi(\bmu)$, respectively.

To find the next sample point $\bmu_{{\gi+1}}$ we selct the parameter value
$\bmu$ for which the RBM Ritz pairs
maximise the residual of the high-dimensional eigenvalue problem, i.e.,
\begin{equation}
    \label{eq:argmax}
    \bmu_{{\gi+1}} 
    =
    \argmax_{\bmu\in\Xitrain} \, \res_\gi(\bmu)
\end{equation}
with
\[
    \res_\gi(\bmu)
    :=
    \sqrt{
    %\left(
    \sum_{i=1}^m
    \left\| \sH(\bmu) \sBn \phirbsi(\bmu) - \lambdarb(\bmu)  \sBn \phirbsi(\bmu)\right\|_{\mathcal N}^2}%\right)^\frac12.
\]
Thanks to the \textit{affine decomposition}, the residuals can be obtained by
\begin{align}
    \nonumber 
    \res_\gi(\bmu)^2
    =&
    \sum_{q,q'=1}^{Q} \theta_q(\bmu)^*\theta_{q'}(\bmu) \, \sum_{i=1}^m\phirbsi(\bmu)^\dagger \,\sHrbqqp \,\phirbsi(\bmu)
    \\&
    \label{eq:Res_aff_3}
    - \lambdarb(\bmu)^2 \sum_{i=1}^m\phirbsi(\bmu)^\dagger \,\sb\,     \phirbsi(\bmu).
\end{align}
Here, the ($d_\gi \times d_\gi$)-matrices $\sHrbqqp:=\sBn^\dagger \sH_q\sH_{q'} \sBn$ are parameter-independent and can be computed once and for all as soon as $\sBn$ is known.

Of note, once the above quantities have been computed and
stored, the computational time required to assemble~\eqref{eq:Hn_assembly},
to obtain a solution of the reduced eigenvalue problem~\eqref{eq:Reduced_evp}
and finally to compute the residual~\eqref{eq:Res_aff_3}
no longer scales with $\mathcal N$, but only depends on $d_\gi$ and $Q$.
Thus, the scan of the training grid via a simple loop is of affordable complexity.

Having determined $\bmu_{{\gi+1}}$ the corresponding high-dimensional
eigenvalue problem~\eqref{eq:Hdim_evp} is solved exactly with a solver of choice.
The resulting $m_{\gi+1}$ ground-states $\bm\Psi(\bmu_{\gi+1})$ are added
to $\V_\gi$ to yield the new space $\V_{\gi+1}$.

Let us make two remarks.
First, the snapshots $\bm\Psi(\bmu_{\gi+1})$ might be highly linearly dependent
to the existing basis $\sBn$.
It is therefore advised to act upon $\bm\Psi(\bmu_{\gi+1})$ in order to reduce
the condition number of the reduced Hamiltonian $\sHrb(\bmu)$, e.g.,
by (approximately) orthogonormalizing them for defining $\sBnp$.
Indeed, by applying a singular value decomposition (SVD) or column-pivoted QR decomposition
to the orthogonal projection onto the complement of $\V_\gi$ one can obtain a reduced
set of vectors
\begin{equation}
    \label{eq:compression}
    \sUn
    =
    \mbox{compress}\big(
    \bm\Psi(\bmu_{\gi+1}) - \sBn^\dagger \sb^{-1} \sBn \bm\Psi(\bmu_{\gi+1}), {\tt tol}\big),
\end{equation}
in which unnecessary modes from $\bm\Psi(\bmu_{\gi+1})$,
which differ less then the target tolerance ${\tt tol}>0$ from $\sBn$,
are dropped.
In the example of the SVD, this consists simply of choosing as $\sUn$ those first left eigenvectors
corresponding to singular values larger than ${\tt tol}$.
From $\sUn$ the basis for step $\gi+1$ is constructed in the usual hierarchical manner
concatenating $\sBnp = [\sBn \,|\, \sUn]$.

Second, let us highlight that we do not actually need the full
$\calN$-dimensional formulation of the full solutions $\bm\Psi(\bmu_\gi)$, but
merely their contractions (scalar products and expectation values) to arrive at
the definition of the (compressed) matrices $\sb$, $\sHrbq$ and $\sHrbqqp$. 
This makes the framework in principle compatible with a Tensor Network based
solver, where $\bm\Psi(\bmu_\gi)$ is obtained in an economic form,
which is polynomially expensive in the number of system
constituents~\cite{Schollwoeck2011,Verstraete2008,Eisert2013,Orus2014,Silvi2019}.

As already emphasized, starting with an initial parameter value
$\bmu_{1}\in\Xitrain$, one repeats this procedure until the maximal residual
error over the training grid is small enough.
This yields an $N$-dimensional reduced basis based on $\nf$ eigenvalue
computations, i.e. $N= d_{\nf}$.
In this manner, the high-dimensional eigenvalue problem~\eqref{eq:Hdim_evp} only needs to be solved $\nf$-times. 
The solution of these $\nf$ problems as well as the compression step~\eqref{eq:compression} and the computation of the reduced matrices $\sHrbq$, $\sHrbqqp$, $\sb$ are the only steps that depend on $\mathcal N$.

Note that the actual surrogate model, i.e., the existing reduced basis
approximation $\Phirb(\bmu) = \sBn \phirb(\bmu)$ which can be cheaply computed
for any $\bmu$, can be used in order to generate accurate initial guesses
$\Phirb(\bmu_{\gi+1})$ for the high-dimensional eigenvalue computation
$\bm\Psi(\bmu_{\gi+1})$. Moreover, the number of points $\nf$ where the greedy
algorithm needs to compute the truth solution is typically much smaller than
the number $M$ of grid-points in $\Xitrain$.

The algorithm is schematically summarized in Algorithm~\ref{alg:offline} where the $\mathcal N$-dependent bottlenecks are marked with a subscript $\mathcal N$.

%%%%%%%%%%%%%%%%%%%%%%
%%%%%%% Offline %%%%%% 
%%%%%%%%%%%%%%%%%%%%%%
\begin{algorithm}[!htb]
  \SetKwData{rbmo}{rbm$_1$}%
  \SetKwData{rbm}{rbm$_n$}%
  \SetKwData{rbmp}{rbm$_{n+1}$}%  
  \SetKwData{rbmN}{rbm$_{\nf}$}%   
  \SetKwData{resn}{res$_n(\bmu)$}%
  \SetKwFunction{compress}{{compress$_{\mathcal N}$}}%
  \SetKwFunction{truth}{{truth-solver$_{\mathcal N}$}}%
  \SetKwFunction{rbsolver}{rb-solver}%  
  \SetKwFunction{residual}{residual}%  
  \SetKwFunction{assemble}{{assemble$_{\mathcal N}$}}%
  
  \BlankLine%
  
  \KwData{training grid $\Xitrain\subset \P$, 
  $\bmu_{1}\in \Xitrain$,
  $\nf$ the number of truth eigenvalue computations
  }
  \KwResult{A surrogate reduced basis model \rbmN$\equiv \{ \sB_{\nf}, \sb, \sHrbq, \sHrbqqp\}$ with $N$ basis functions based on $\nf$ truth solves.}

  \BlankLine%

    \BlankLine%

    $\bm\Psi(\bmu_{1})$ \( \leftarrow \) 
    \truth{$\bmu_{1}$}
    \hfill \eqref{eq:Hdim_evp}
    
    \rbmo \( \leftarrow \) \compress{$\bm\Psi(\bmu_{1})$} 
    \hfill \eqref{eq:compression}
    
    \BlankLine%
    \BlankLine%
    
    \While{$\max_{\bmu\in\Xitrain} $ \, \resn $>{\tt tol}$}{
        \BlankLine%
        \For{$\bmu\in\Xitrain$}{
            $\Phirb(\bmu)$ \( \leftarrow \) 
            \rbsolver{\rbm}
            \hfill \eqref{eq:Reduced_evp}
            
            \resn \( \leftarrow \) 
            \residual{$\Phirb(\bmu)$,\rbm}
            \hfill \eqref{eq:Res_aff_3} 
        }
        \BlankLine%
        
        $\bmu_{{n+1}}$
        \( \leftarrow \) 
        $\argmax_{\bmu\in\Xitrain}$ \, \resn
        \hfill \eqref{eq:argmax}
        
        $\bm\Psi(\bmu_{{n+1}})$ \( \leftarrow \) 
        \truth{$\bmu_{{n+1}}$}
        \hfill \eqref{eq:Hdim_evp}
    
        $\sUn$ 
        \( \leftarrow \) 
        \compress{$\sBn,\bm\Psi(\bmu_{{n+1}})$} 
        \hfill \eqref{eq:compression}
        
        \rbmp
        $\equiv \{ \sB_{\nf+1}, \sb, \sHrbq, \sHrbqqp\}$ 
        \ldots
        
        \hfill
        \( \leftarrow \) 
        \assemble{$\sBn,\sUn$}

    }
    precompute $\sOrbr = \sB_N^\dagger \sO_r \sB_N$
    
    \BlankLine%
  \caption{Overview of the offline step.\label{alg:offline}}
\end{algorithm}

\paragraph{Online:}
Once the reduced basis is assembled, the so-called online step can be started.
It is the evaluation of the map $\bmu \mapsto \phirbN(\bmu)$ for \textit{any}
$\bmu\in\P$ where $\phirbN(\bmu)$ denotes the solution
to~\eqref{eq:Reduced_evp} and which is \textit{independent} of $\mathcal N$
(only depends on $N$ and $Q$).
Note that we drop in the online-step the superscript $\Ngi$ for all relevant
quantities since the reduced basis is fixed at this point.

From the RBM a representation of the solution in the high-dimensional Hilbert space
$\C^{\calN}$ can then be obtained by $\sB_N \phirbN(\bmu)$, which scales
linearly in $\mathcal N$. However, typically this is not needed for any practical
purpose --- which is one of the strengths of the RBM-framework.
Most importantly
the computation of an output functional can indeed be performed in a
complexity independent of $\calN$ using
\begin{equation}
    \label{eq:online_output_fcl}
    \cOrb(\bmu;p) = \sum_{r=1}^{R} \alpha_r(\bmu;p) \sum_{i=1}^m \phirbsNi(\bmu)^\dagger \sOrbr \phirbsNi(\bmu)
\end{equation}
with precomputed $\sOrbr = \sB_N^\dagger \sOr \sB_N$.
In turn, the output evaluation $\bmu \mapsto \cOrb(\bmu;p)$ becomes
\textit{independent} of $\mathcal N$ and only depends on $N$, $Q$ and $R$.

The algorithm is schematically summarized in Algorithm~\ref{alg:online}.

%%%%%%%%%%%%%%%%%%%%%%
%%%%%%% Online %%%%%%% 
%%%%%%%%%%%%%%%%%%%%%%
\begin{algorithm}[!htb]
  \SetKwData{rbmo}{rbm$_1$}%
  \SetKwData{rbm}{rbm$_n$}%
  \SetKwData{rbmp}{rbm$_{n+1}$}%  
  \SetKwData{rbmN}{rbm$_{\nf}$}%   
  \SetKwData{resn}{res$_n(\bmu)$}%
%   \SetKwData{sol}{sol}%
  \SetKwFunction{rboutput}{rb-output}%
  \SetKwFunction{compress}{compress}%
  \SetKwFunction{truth}{truth-solver}%
  \SetKwFunction{rbsolver}{rb-solver}%  
  \SetKwFunction{residual}{residual}%  
  
   \BlankLine%
   
  \KwData{\rbmN, $\sOrbr$, $\bmu\in\P$ 
  }
  \KwResult{$\lambdarbN(\bmu)$, $\cOrb(\bmu;p)$}

  \BlankLine%

  \BlankLine%

    \BlankLine%

            $\lambdarbN(\bmu), \phirbN(\bmu)$ \( \leftarrow \) 
            \rbsolver{\rbmN}
            \hfill \eqref{eq:Reduced_evp}
            
            $\cOrb(\bmu;p)$
            \( \leftarrow \) 
            \rboutput{$\phirbN(\bmu)$}
             \hfill \eqref{eq:online_output_fcl}
    
     \BlankLine%
      \caption{Overview of the online step.\label{alg:online}}
\end{algorithm}

\section{Results}
\label{sec:Results}
We test the methodology on the two example applications introduced in Sections~\ref{ssec:rydberg} and~\ref{ssec:triangle}. 
The former is interesting as it allows to consider different chains of variable lengths and analyze the structure factors. 
The latter is interesting since it contains degeneracies due to symmetries allowing to test the methodology in this case. 

To conduct our bench-marking tests,
we implemented the method, as outlined above, in Julia~\cite{Bezanson2017}.
For our computation we exploit the sparsity of the Hamiltonians $\sH(\bmu)$
using compressed sparse column matrix storage. The exact ground states have been obtained
iteratively using the locally optimal preconditioned conjugate gradient~(LOBPCG)
algorithm~\cite{Knyazev2001,Hetmaniuk2006,Duersch2018}
as implemented in the density-functional toolkit~(DFTK)~\cite{DFTK,DFTKjcon}
using an incomplete (sparse) Cholesky factorization as a preconditioner.
Whenever a truth-solve needs to be performed and a reduced basis
is already existing, we use the RBM-approximation as a surrogate to provide
an accurate initial guess.
We ensure the right number of degenerate ground states is found by adapting
the number of targeted eigenpairs during the iterative diagonalization until the
obtained eigenvalue gap is larger than the convergence tolerance.
Our scheme has been carefully verified to obtain the correct number of eigenpairs
even in the highly degenerate cases of the antiferromagnetic spin-$\frac12$ triangles.
Notice that for the chain of Rydberg atoms the LOBPCG sometimes fails to converge
within a given number of iterations.
In such cases we fall back to a
full diagonalization of the Hamiltonian using a direct method.
This happens especially frequently in the interstitial region between two clear phases.
Overcoming this limitation, e.g.~by including not only the ground state but also
the closest manifolds of excited states in the reduced basis,
is an interesting direction for future work.

In the upcoming analyses, we will use
\begin{align}
    \label{eq:valerror}
    \valerror &= \max_{\bm\mu\in\Xitest}
    \frac{| \lambda(\bm\mu) -\lambdarbN(\bm\mu) |}{|\lambda(\bm\mu)|}.
\end{align}
in order to asses the quality of the reduced basis approximation
to the eigenvalue over a given test grid $\Xitest$ that is different from the training
grid $\Xitrain$.
The computation of the eigenvector error is slightly more complicated with
degenerate ground-states. For that purpose we will compare the spectral
projectors onto the different eigenspaces given by the density matrices, i.e.,
we will consider the following error quantity
\begin{align}
    \label{eq:vecerror}
    \vecerror &= \max_{\bm\mu\in\Xitest}
    \frac{
    \| \bm\Psi(\bm\mu)\bm\Psi(\bm\mu)^\dagger -\PhirbN(\bm\mu) \PhirbN(\bm\mu)^\dagger\|_{F
    }}{\| \bm\Psi(\bm\mu)\bm\Psi(\bm\mu)^\dagger\|_{F
    }}
    ,
\end{align}
where $\PhirbN(\bm\mu) = \sB_{\nf} \phirbN(\bmu)$, the Ritz vector(s), and $\|\cdot\|_F$ denotes the Frobenius norm.
Further, the output functional will be measured as
\begin{equation}
    \label{eq:sferror}
    \sferror = \max_{\bm\mu\in\Xitest}
    \frac{ \| \mathcal S(\bmu ;\,\cdot\,) - \mathcal S_{\tt rb}(\bmu ;\,\cdot\,)\|_{F}}{\| \mathcal S(\bmu ;\,\cdot\,)\|_{F}},
\end{equation}
where the truth structure factor $\mathcal S$ and its reduced basis approximation $\mathcal S_{\tt rb}$ are evaluated on a fine enough grid in Fourier space.

Note that we use equispaced training and test grids. As an alternative, one can
use random training grids, such as proposed in~\cite{hesthaven2014efficient}
for greedy-strategies, but one may not discover special cases and degeneracies
arising at very particular parameter values, such as diagonals in parameter space etc.

In both our test cases, we investigate the properties of the greedy algorithm, analyze the convergence of the reduced basis approximation with respect to the basis size $N$ and compare it with the optimal decay revealed by the singular values of the snapshot matrix discussed at the beginning of this section.
Let us emphasize that the greedy algorithm does not require the truth-solution at all grid-points, we only use them here in order to measure the error in different metrics to assess the quality of the greedy algorithm, but this is a truly academic investigation.
Furthermore, we will present results on the output functional in form of the appropriate structure factors as defined above. 

%%%%%%%%%%%%%%%%%%%%%%%%%%%%%%%
%%%%%%% RYDBERG MODEL %%%%%%%%%
%%%%%%%%%%%%%%%%%%%%%%%%%%%%%%%
\subsection{Chain of excited Rydberg atoms}
\begin{figure}[t!]
    \centering
    \includegraphics[width=0.45\textwidth]{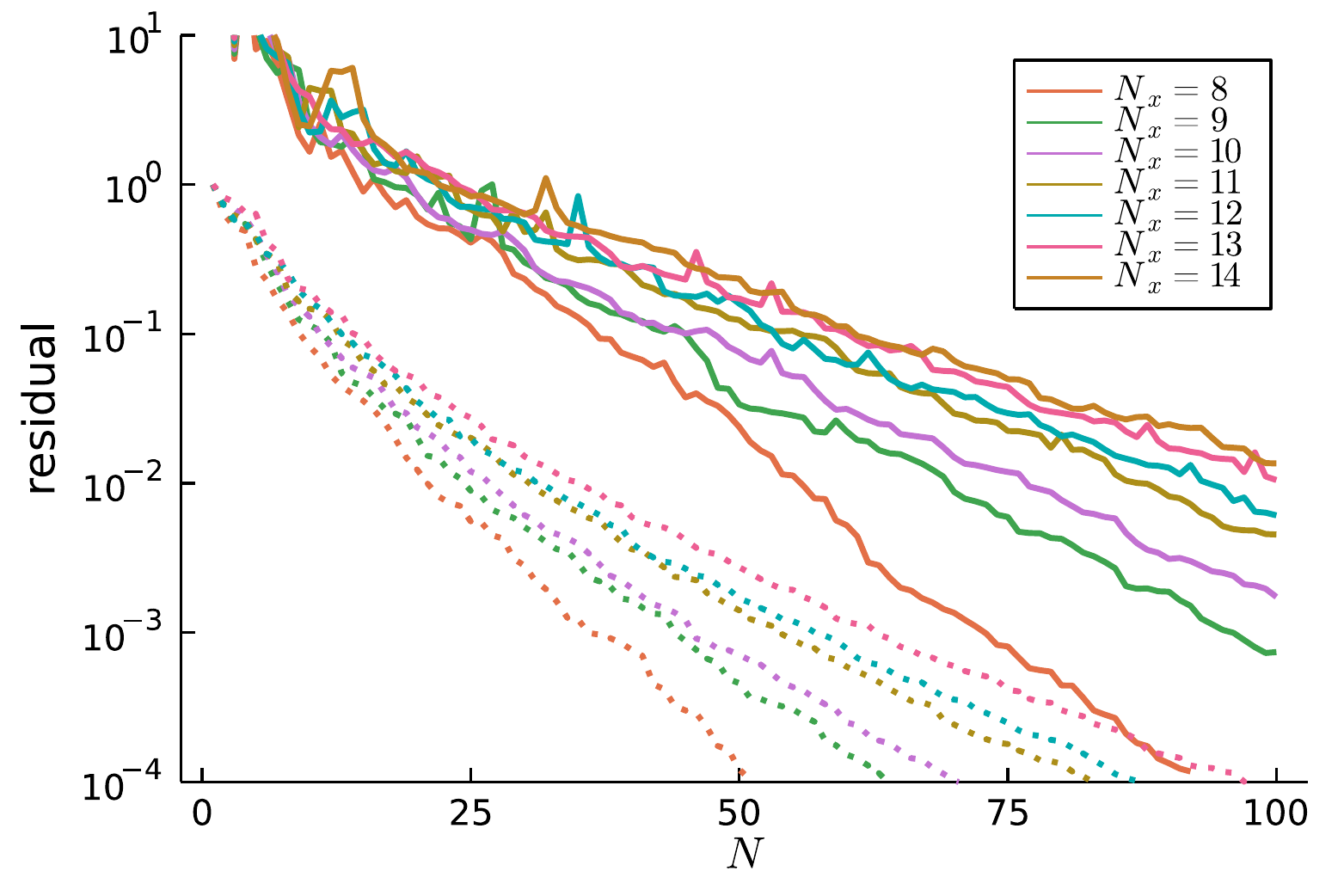}
    \caption{
    Decay of the maximal residual over the training grid $\Xitrain$ during the greedy algorithm with respect to $N$ for chains of Rydberg atoms of different lengths $\nx$. The singular values are dotted as a reference.
    }
    \label{fig:Ryd_res}
\end{figure}

Here, we consider a chain of excited Rydberg atoms of varying size $\nx$ as in Eq.~\eqref{eq:HamRydberg} and a parameter domain $\P=[0, 5]\times[0.5, 4]$,
% \SW{can we motivate this, e.g., by giving a reference?}
where we recall that $\mu_1=\Delta/\Omega$ and $\mu_2=n_S$.
We apply the greedy-sampling strategy outlined in Section~\ref{sec:Method} using a training grid $\Xitrain$ consisting of $50\times 50$ uniformly spaced points in $\P$. 
The evolution of the maximal residual over the training grid $\Xitrain$ during the offline-step is illustrated in Figure~\ref{fig:Ryd_res} where we also show the decay of the singular values. 
We observe that the decay of the residual nicely follows the decay of the singular values up to a constant offset. As was already observed for the decay of the singular values, the effective dimension $N$ for fixed tolerance only increases mildly for increasing~$\nx$.

\begin{figure}[t!]
    \centering
    \includegraphics[width=0.45\textwidth]{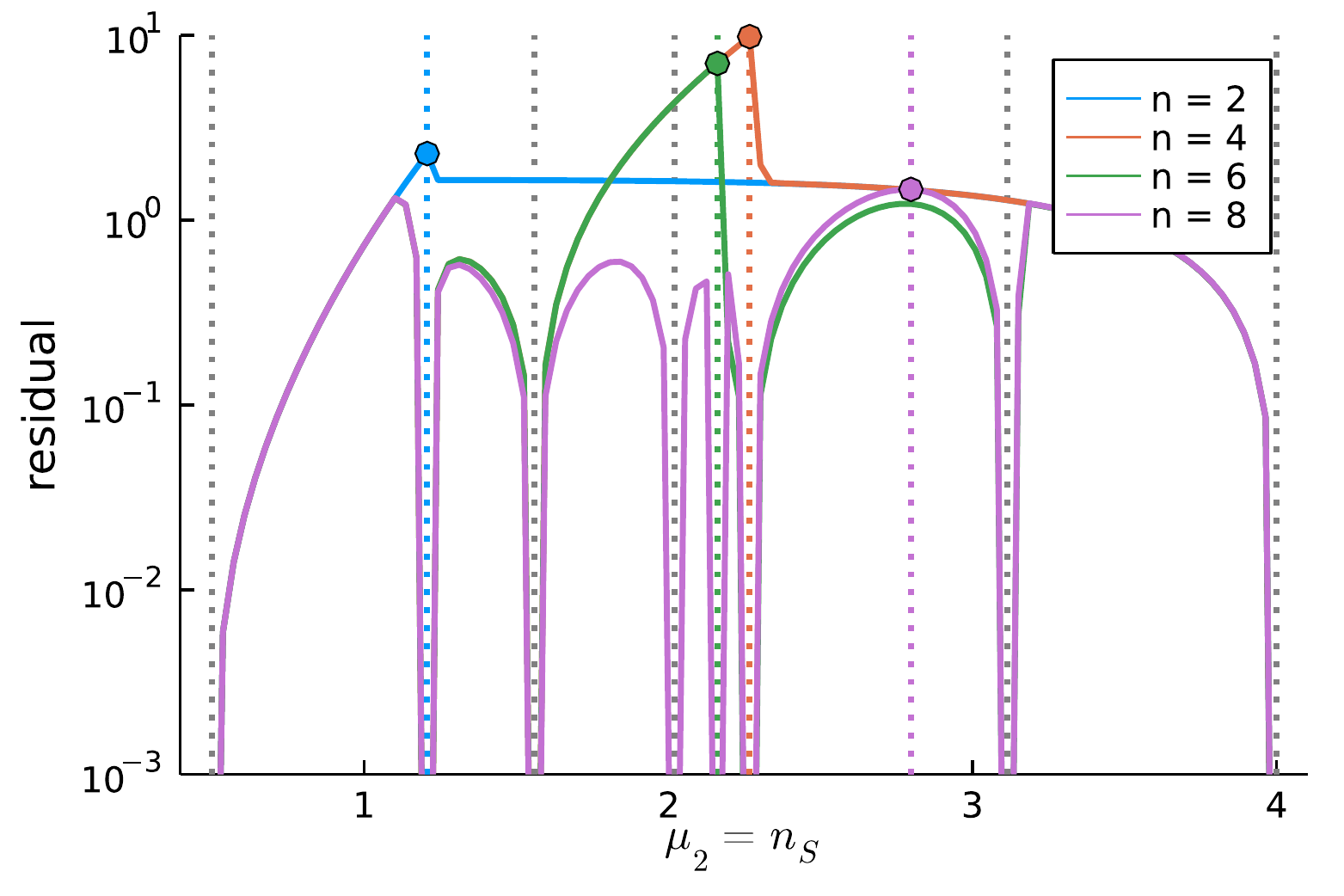}
    \caption{
    Illustration of the greedy algorithm for a one-dimensional parameter space $\{4.5\}\times [0.5,4]$ in the case of a chain of Rydberg atoms with $\nx=13$.
    }
    \label{fig:Ryd_educative}
\end{figure}

\begin{figure}[t!]
    \centering
    \begin{overpic}[width=0.45\textwidth]{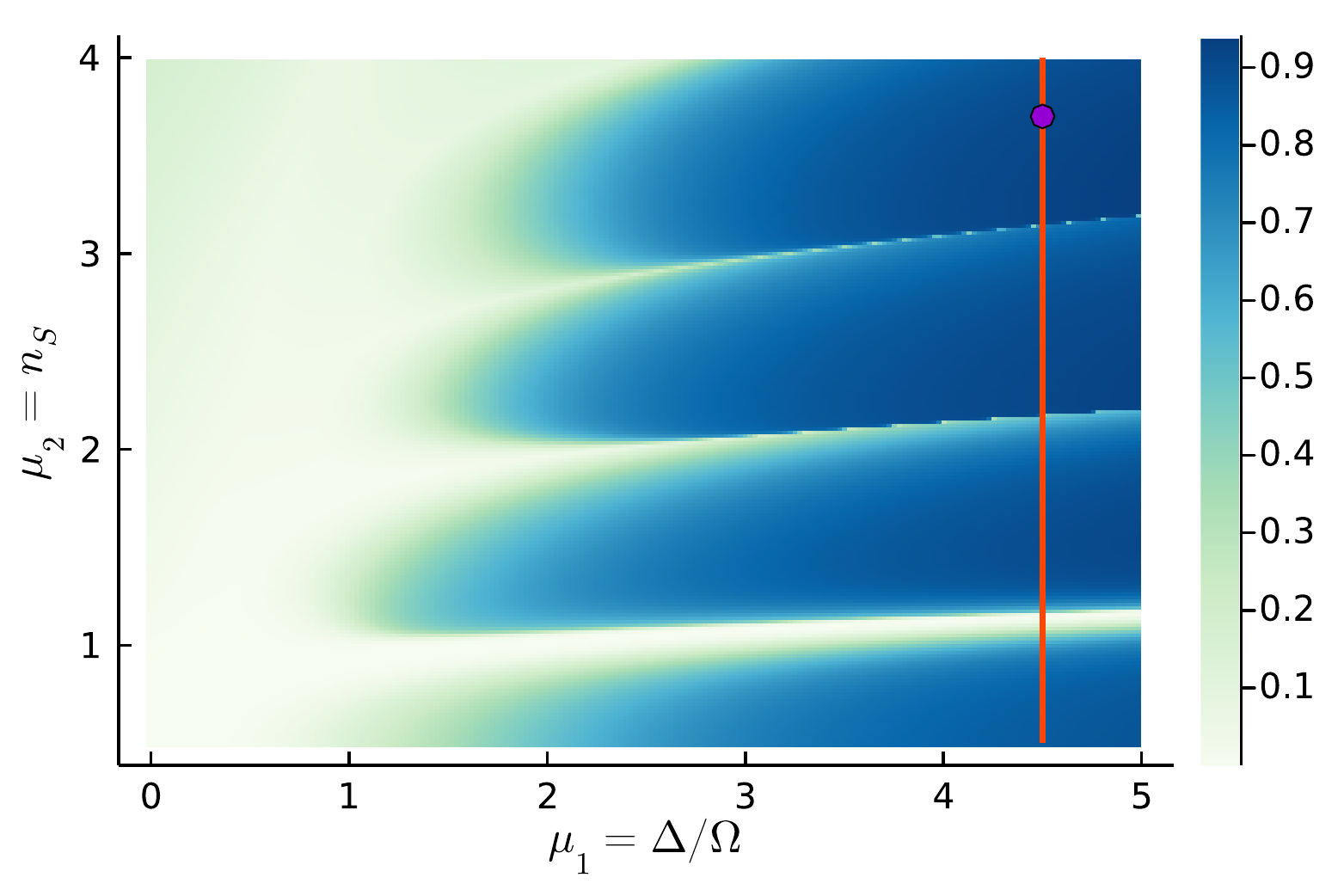}
    \put (0,60) {(a)}
    \end{overpic}
    \begin{overpic}[width=0.45\textwidth]{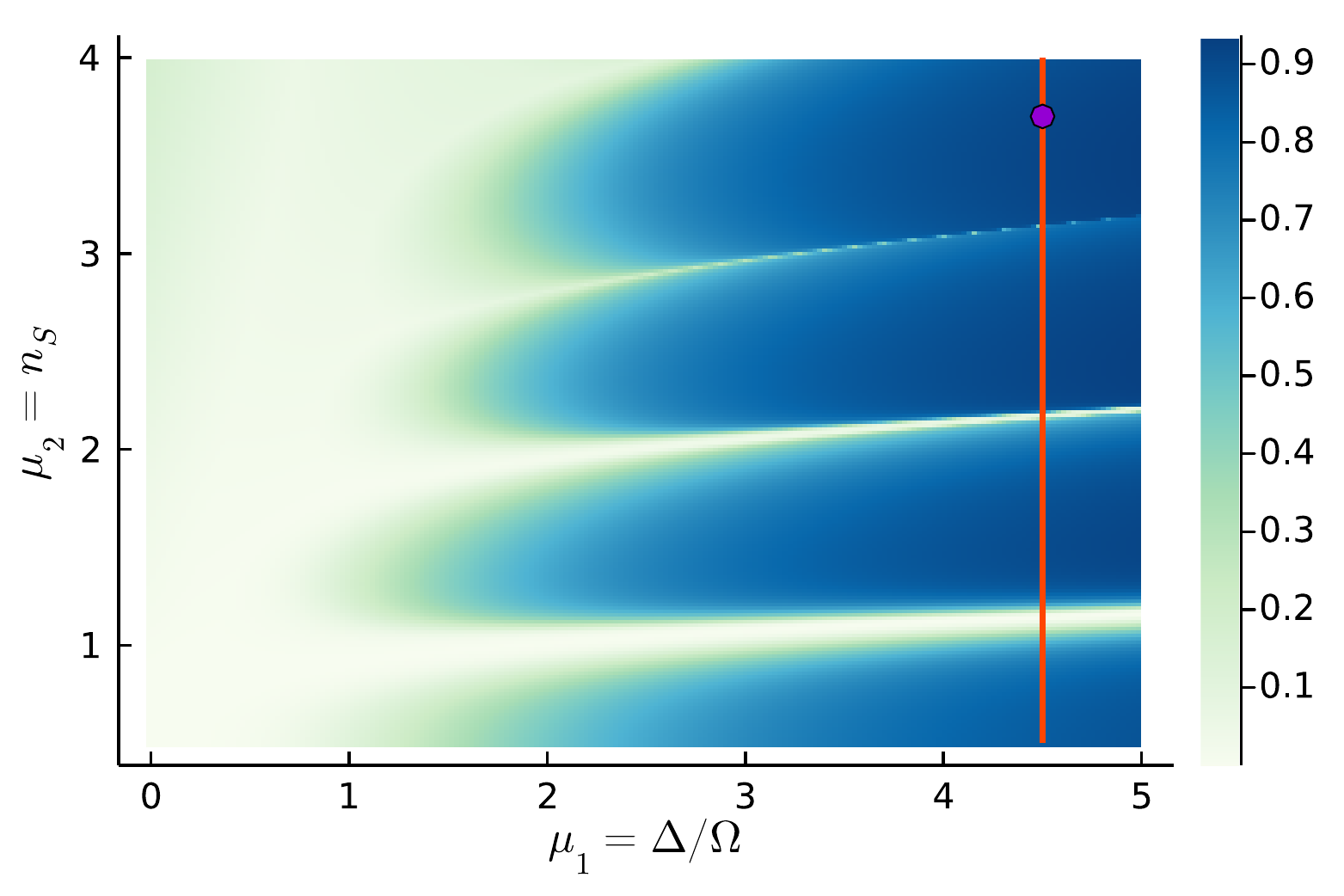}
    \put (0,60) {(b)}
    \end{overpic}
    \caption{
    The occupation number using the RBM with $N=20$ (a) and $N=100$ (b) over the parameter domain for a chain of Rydberg atoms with $\nx=13$.
    The orange line at $\mu_1 = 4.5$ indicates the direction along which the structure factor is plotted in Figure \ref{fig:Ryd_sf_map}
    and the purple dot denotes the value at which its convergence is discussed in Figure \ref{fig:Ryd_sf_modes}.
    }
    \label{fig:Ryd_occ_map}
\end{figure}

In order to give an illustration how the greedy acts, we show in Figure~\ref{fig:Ryd_educative} the error profiles for different values of $N$ for the one-dimensional parameter space along the line of fixed $\mu_1=4.5$ (corresponding to the orange line in the upcoming Figure~\ref{fig:Ryd_occ_map}).
We illustrate the profile of the residual over the one-dimensional parameter space during the greedy iterations, i.e., for different values of $n=2,4,6,8$. The maximum of each curve is marked by a dot and the first eight sample points $\bmu_n$ ($n=1,\ldots,8$) by vertical lines. We observe that in agreement with theory, at each iteration $n$, the residuals vanish at the sample points $\bmu_i$, for $i=1,\ldots,n-1$, since the truth solution at those points already belongs to the reduced basis space. By always adding the worst approximation (as quantified by the residual), this approach allows us to reduce the maximum residual over the parameter space and over the iterations $n$, though not necessarily in a monotonic fashion (see also Figure~\ref{fig:Ryd_res}). 

% \BS{More explanations needed how the greedy works}

%%%%%% Rydberg: residual decay %%%%%% 
%%%%%% end figure %%%%%% 

%%%%%% Rydberg: educative example greedy %%%%%% 
%%%%%% end figure %%%%%% 

%%%%%% Rydberg: error decay %%%%%% 
\begin{figure}[t!]
    \centering
    \includegraphics[width=0.45\textwidth]{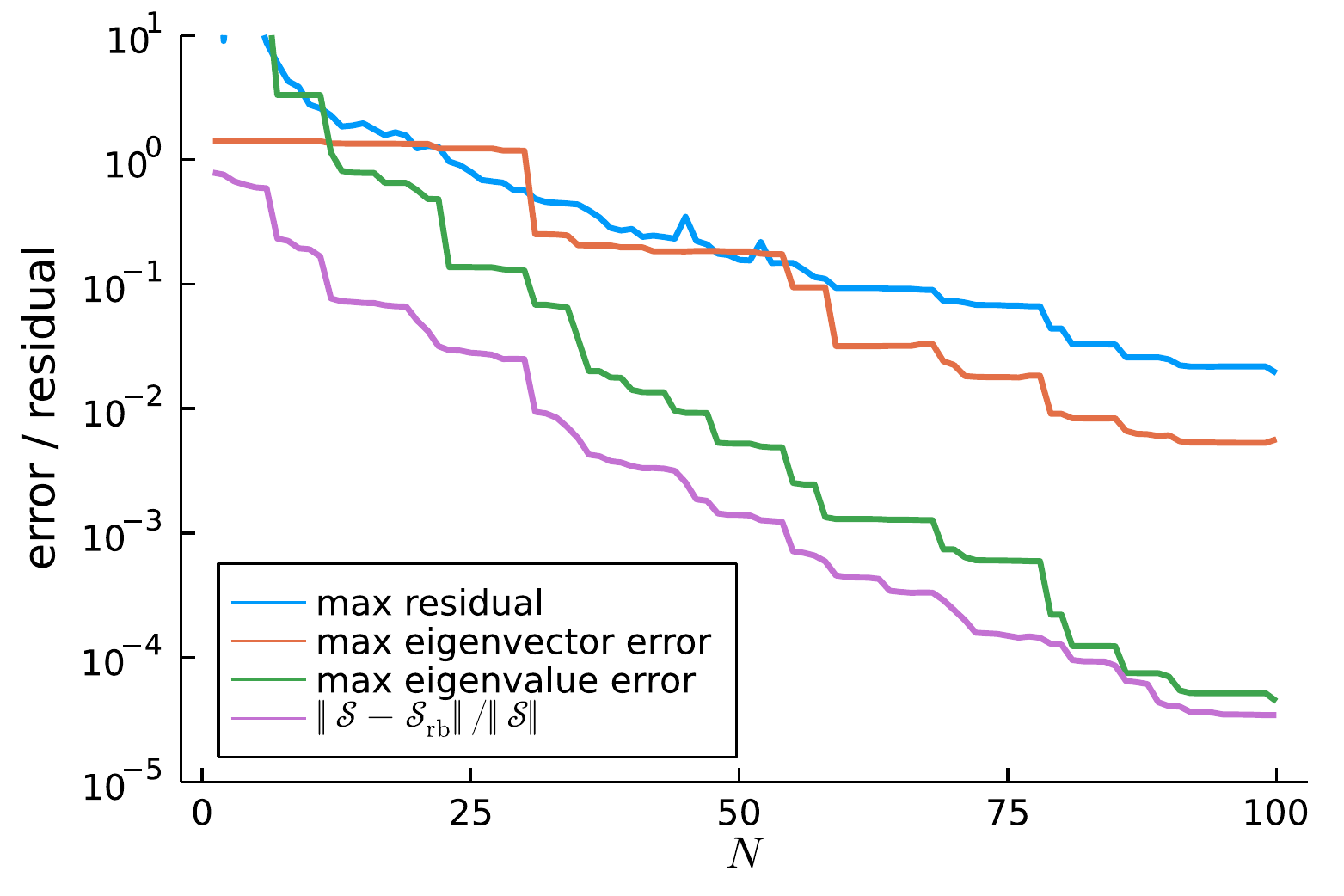}
    \caption{
    Decay of the error quantities of the RBM and the residual during the greedy algorithm with respect to $N$ for a chain of Rydberg atoms with $\nx=13$.
    }
    \label{fig:Ryd_error}
\end{figure}
%%%%%% end figure %%%%%% 
%%%%%% Rydberg: error map %%%%%% 
\begin{figure}
    \centering
    \begin{overpic}[width=0.45\textwidth]{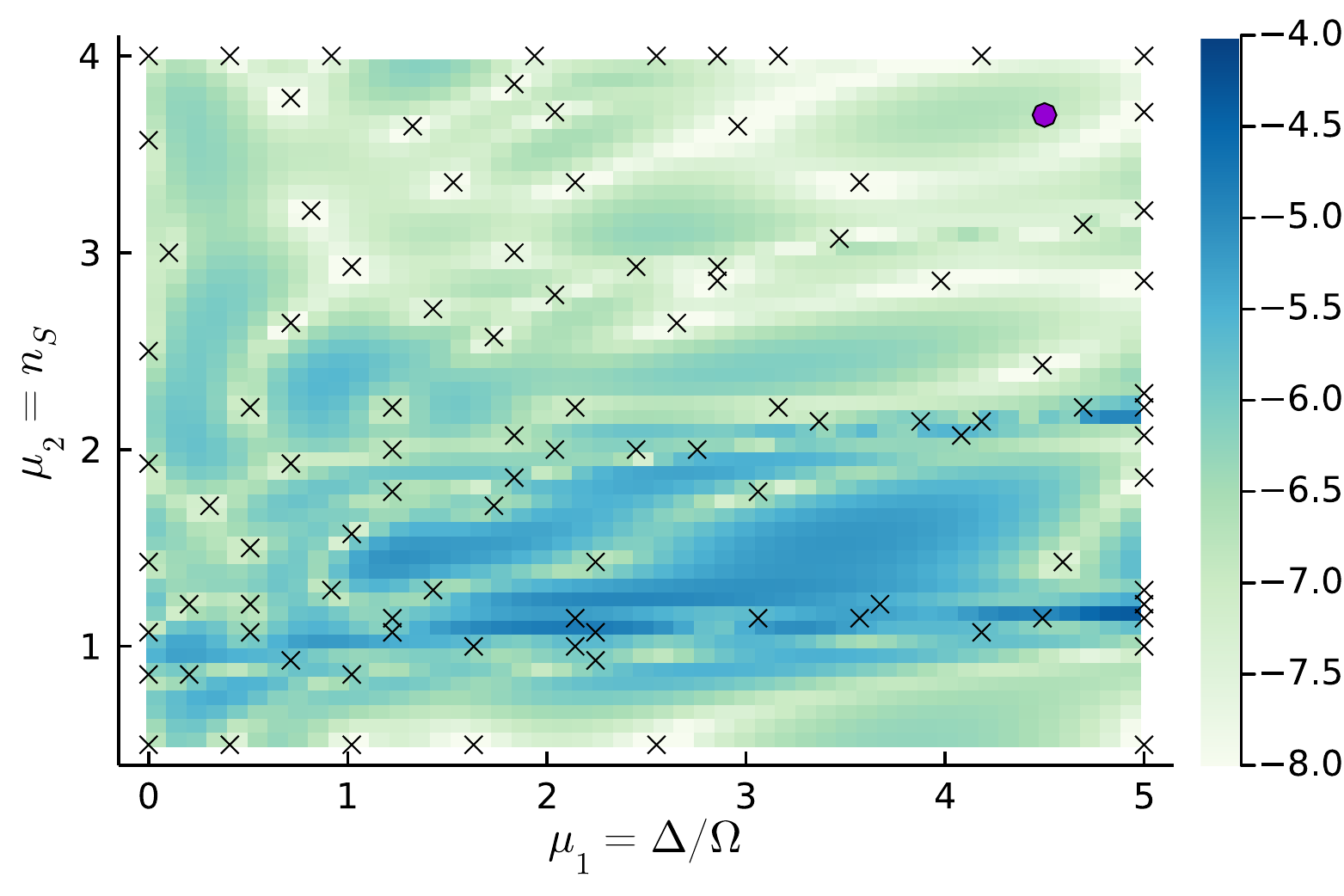}
    \put (0,60) {(a)}
    \put (91,4) {$\log_{10}$}
    \end{overpic}
    \begin{overpic}[width=0.45\textwidth]{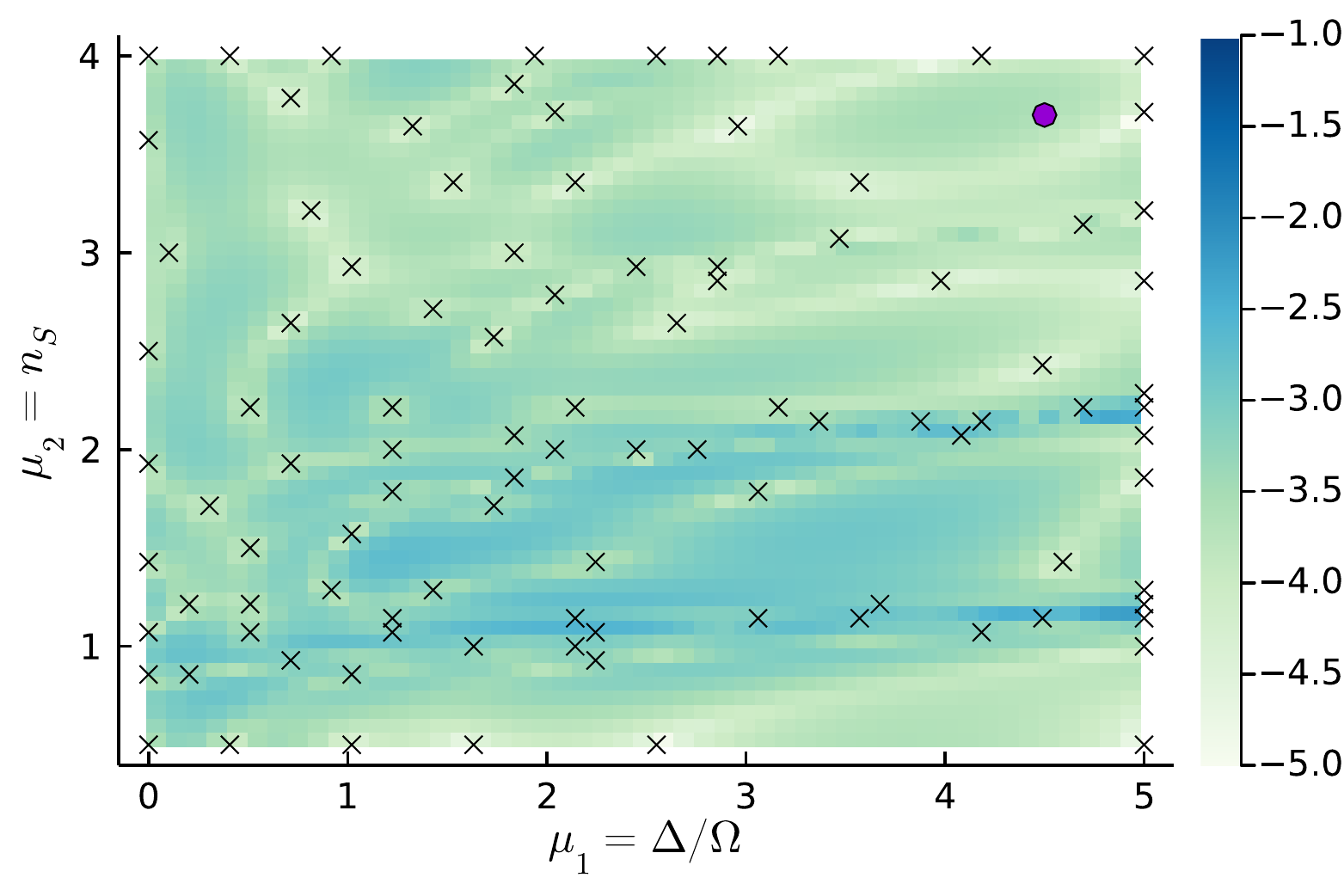}
    \put (0,60) {(b)}
    \put (91,4) {$\log_{10}$}
    \end{overpic}
    \caption{
        Logarithmic distribution of the eigenvalue error (a) and
        eigenvector error (b) of the RBM for $N=100$
        over the parameter domain and the sample points selected by the greedy
        (crosses) for a chain of Rydberg atoms with $\nx=13$. The purple dot
        locates a specific parameter set that is examined in more detail in later
        figures.
    }
    \label{fig:Ryd_error_map}
\end{figure}
%%%%%% end figure %%%%%% 

Returning  to the two-dimensional parameter domain, we test in Figure~\ref{fig:Ryd_error} the actual errors $\valerror$, $\vecerror$ and $\sferror$ of the eigenvalue, eigenvector and structure factors respectively for an increasing sequence of $N$ for fixed $\nx=13$. For testing, we use a test grid $\Xitest$ consisting of $49\times 49$ uniformly space points in $\P$ that are different to the those in the training grid $\Xitrain$.
For comparison we also illustrate the evolution of the residual.
We observe that the error surrogate given by the residual follows closely the real error in the eigenvector and an accelerated convergence of the eigenvalues, which is typical and expected for variational eigenvalue approximations.
The fact that the eigenvector error, the residual and the singular values exhibit the same decay rate means that the greedy strategy assembles the reduced basis in a quasi-optimal fashion in this test case.
Figure~\ref{fig:Ryd_error_map} shows the distribution of the eigenvalue error and eigenvector
error over the test-grid $\Xitest$ for $N=100$ and $\nx=13$ as well as the sample points selected by the greedy algorithm.
Both  eigenvalue and eigenvector errors show very similar behavior
with roughly a constant offset in the error values.
The larger errors in the eigenvectors are to be expected due to the faster decay
in the eigenvalue error apparent in Figure~\ref{fig:Ryd_error}.

Furthermore, Figure~\ref{fig:Ryd_occ_map} shows the highest occupation number of the states in the canonical basis for the same parameters as considered above for $N=20$ and $N=100$.
In the eyeball-norm, one can only detect small differences, for instance in the transition between the $\mathbb Z_2$ and $\mathbb Z_3$ crystalline phases.

\begin{figure}[t!]
    \centering
    \includegraphics[width=0.45\textwidth]{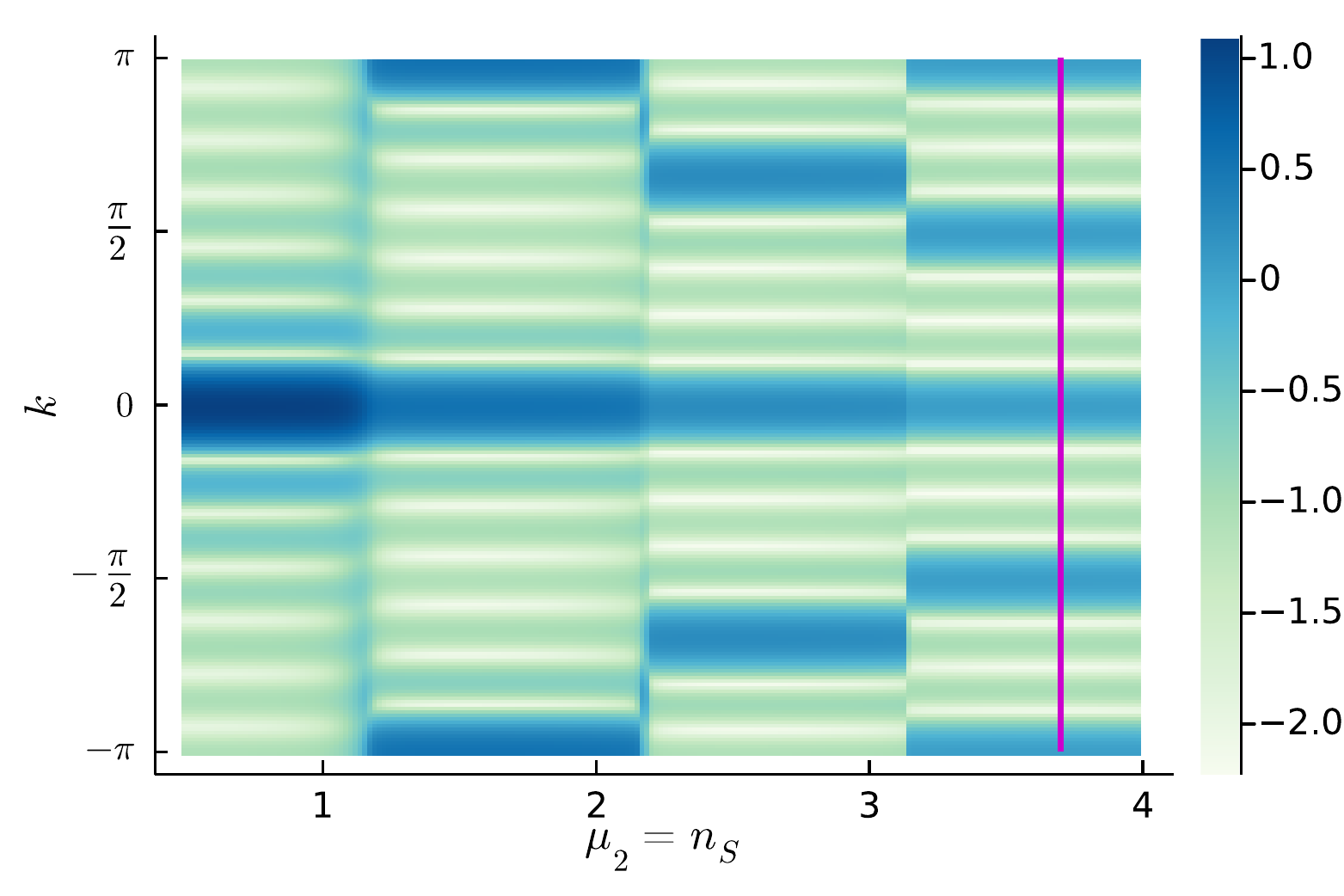}
    \caption{
    The structure factor (in log-scale) using the RBM as a function of
    $\mu_2=n_S$ for fixed $\mu_1=4.5$ and the wave-number $k$
    for a chain of Rydberg atoms with $\nx=13$.
    The purple line indicates the parameter value at which the convergence  of the structure factor is discussed in Figure \ref{fig:Ryd_sf_modes}.
    }
    \label{fig:Ryd_sf_map}
\end{figure}

In order to identify the dominant modes, we illustrate in Figure~\ref{fig:Ryd_sf_map} the structure factors along the vertical line with $\mu_1=\Delta/\Omega=4.5$ for $N=100$.
We observe that the RBM can capture faithfully the distinct arrangements of Rydberg excitations inside the different lobes, namely their $\mathbb Z_{\ell}$ character with $\ell$ a divisor of $\nx-1$.
Certainly, given the number of Rydberg atoms that we consider here, one cannot expect to resolve even further details about the phase diagram such as the nature of the narrow floating phase between the lobes or the nature of the quantum phase transitions in this model~\cite{Fendley2004CDW,Pohl2010Crystal,rader2019floating,chepiga2021lifshitz,chepiga2021kibble,keesling2019kibble}.

\begin{figure}[t!]
    \centering
    \includegraphics[width=0.45\textwidth]{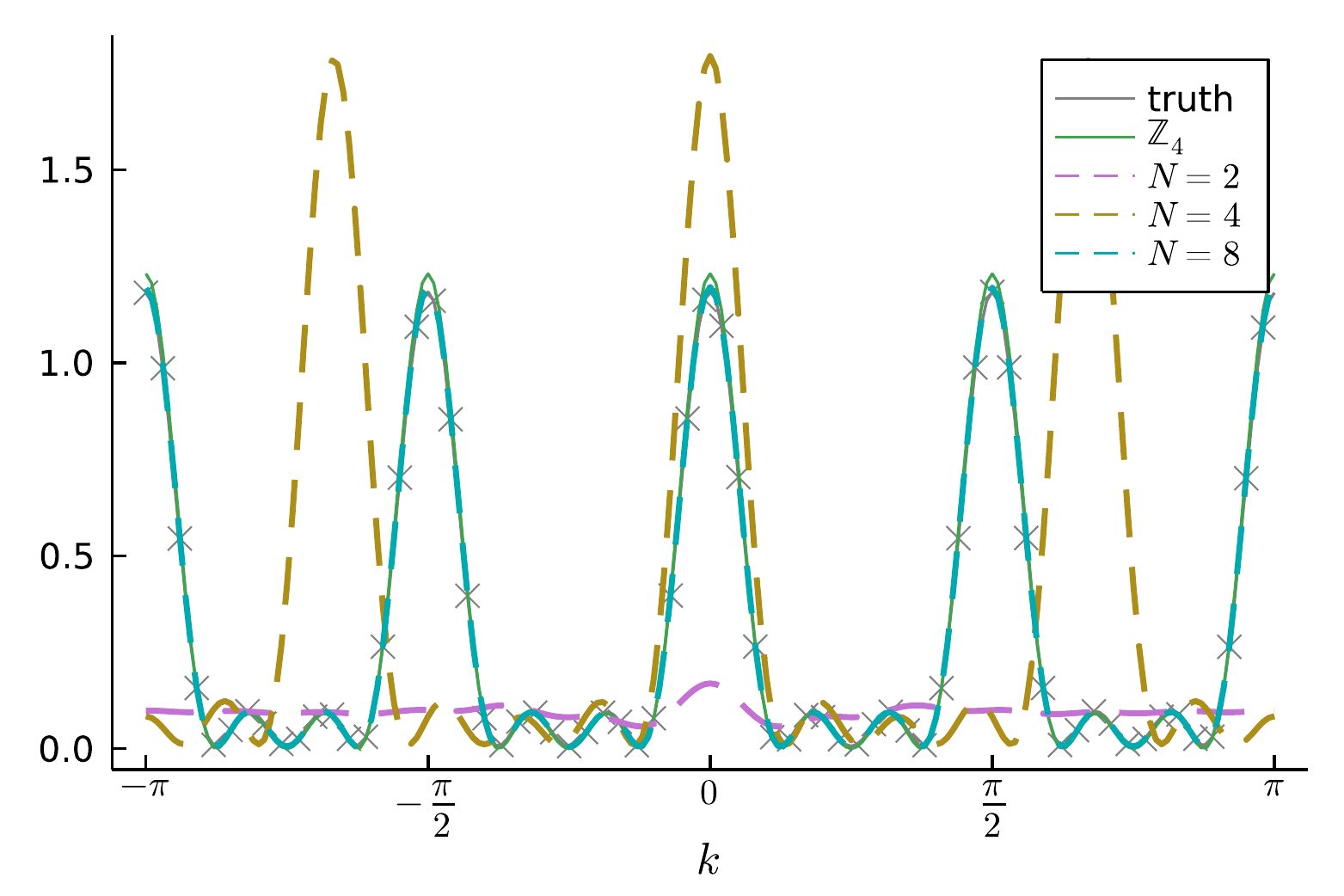}
    \caption{
    The structure factor (in log-scale) using the RBM for different values of $N$
    and fixed $\bmu=(4.5,3.7)$  for a chain of Rydberg atoms with $\nx=13$.
    For sake of completeness, we also illustrate the solution of the truth
    (exact) model and the structure factor of the pure $\mathbb Z_4$-mode.
    Grey crosses are used to highlight the position of the truth curve,
    which is almost perfectly overlaid by the result obtained in the $N=8$ basis.
    }
    \label{fig:Ryd_sf_modes}
\end{figure}

In Figure~\ref{fig:Ryd_sf_modes} we show the convergence of the structure
factor for fixed $\bmu=({4.5, 3.7})$ (corresponding to the purple point in
Figures~\ref{fig:Ryd_error_map} and~\ref{fig:Ryd_occ_map} and the vertical purple line in
Figure~\ref{fig:Ryd_sf_map}) for different values of $N=2,4,8$ besides
the structure factors of the pure state $\mathbb Z_4$.
This point $\bmu$ was chosen such that it is not close to any
sampling point of the greedy assembly, see Figure~\ref{fig:Ryd_error_map}.
We observe a very fast convergence in the eyeball-norm.

%%%%%% Rydberg: occupation map %%%%%% 
% \begin{figure}[t!]
%     \centering
%     \includegraphics[width=0.45\textwidth]{rydberg_occmap_20.pdf}
%     \caption{
%     The occupation number using the RBM with $N=20$
%     over the parameter domain for a chain of Rydberg atoms with $\nx=13$.
%     }
%     \MFH{Mention in text}
%     \label{fig:Ryd_occ_map10}
% \end{figure}
%%%%%% end figure %%%%%% 
%%%%%% Rydberg: SF map %%%%%% 
%%%%%% end figure %%%%%% 
%%%%%% Rydberg: SF modes %%%%%% 
%%%%%% end figure %%%%%% 

%%%%%%%%%%%%%%%%%%%%%%%%%%%%%%%%
%%%%%%% TRIANGLE MODEL %%%%%%%%%
%%%%%%%%%%%%%%%%%%%%%%%%%%%%%%%%
\subsection{Antiferromagnetic spin-$\tfrac12$ triangles in \lacumo}
We consider the antiferromagnetic lattice model  of Eq.~\eqref{eq:AFTriangles} with varying sizes $\nx$, $\ny$ and a parameter domain 
\[
    \P=[0, 2]\times[0, 2]\times [0.01,0.1],
\]
where we recall that $\mu_1=J_1/J_3$, $\mu_2=J_2/J_3$ and $\mu_3=J'$. 
In this regime, 
the model constitutes an interesting stress-test for our method, since it exhibits
(for small $J'$)
ground-state manifolds with a degree of degeneracy that varies across the parameter domain. More specifically, we find numerically the maximal degree of degeneracy itself to increase
with $\nx$, $\ny$ as $2^{\nx\ny}$.
Another reason why we focus on the above parameter regime is that this region was previously considered for the compound  \lacumo.

\begin{figure}[t!]
    \centering
    \includegraphics[width=0.45\textwidth]{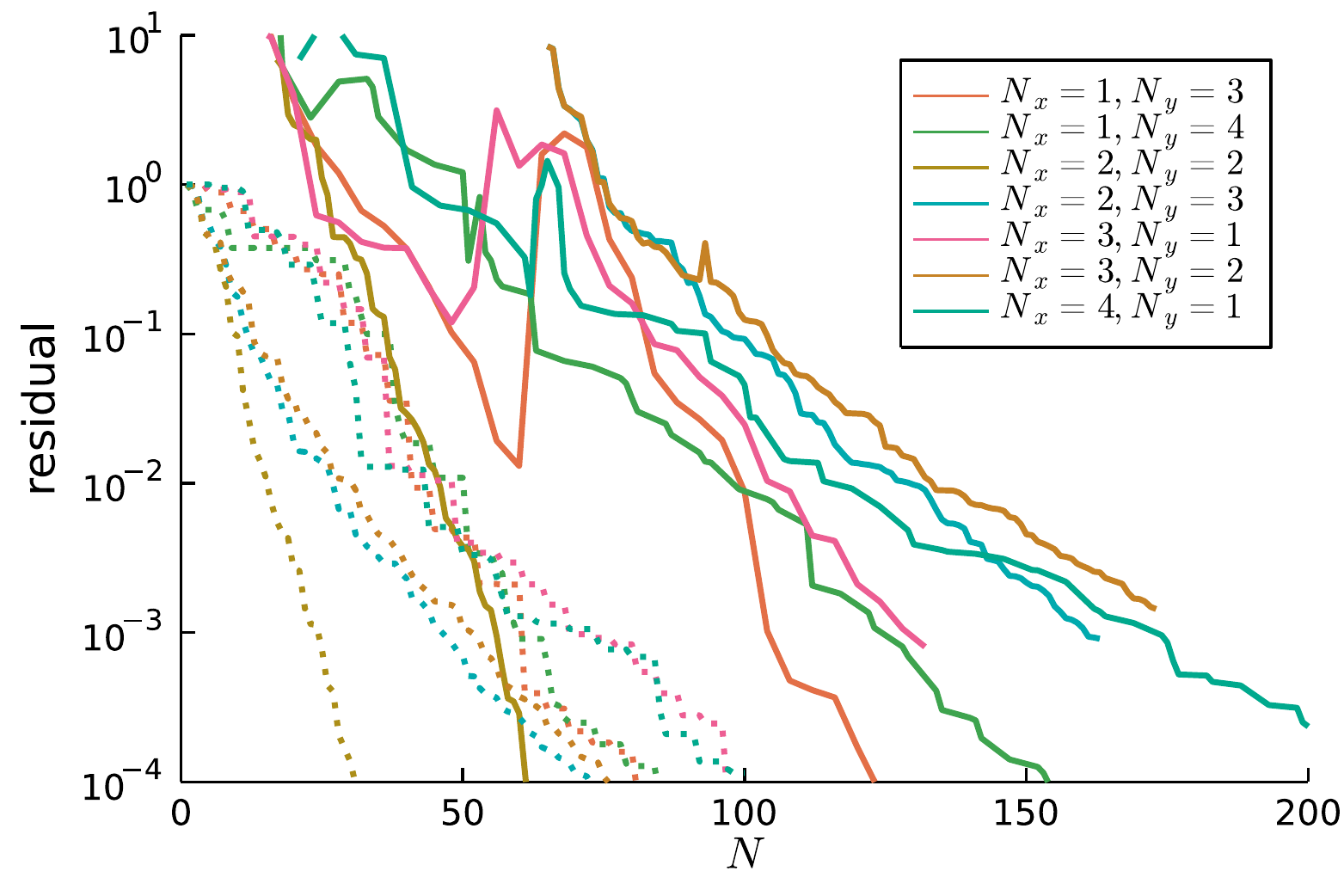}
    \caption{
    Decay of the residual during the greedy algorithm with respect to $N$ for
    the triangle models of different lengths $\nx,\ny$.
    A uniformly spaced grid of $10\times 10\times 10$ points was used
    for training.
    The singular values are dotted as a reference.
    }
    \label{fig:Triangle_res}
\end{figure}

Again, we apply the greedy-sampling strategy outlined in
Section~\ref{sec:Method} first using a training grid $\Xitrain$ consisting of
$10\times 10 \times 10$ uniformly spaced points in $\P$.
The evolution of the residual during the offline-step is illustrated in
Figure~\ref{fig:Triangle_res} where we also show the decay of the singular
values as reference. Also in this case, we observe that the decay of the
residual nicely follows the decay of the singular values.
Ensured by this observation we switch to a larger training grid
$\Xitrain$ of $20\times 20 \times 20$ uniformly spaced points in $\P$
in the remainder of this section.
Note, that a ground truth computation on all 8\,000 parameter values
is computationally very demanding for the larger triangle systems.
For this reason an investigation of the singular value decay as well as
the eigenvalue error is not attempted here.

%%%%%% Triangle: residual decay %%%%%% 
%%%%%% end figure %%%%%% 

\begin{figure}[t!]
    \centering
    \includegraphics[width=0.45\textwidth]{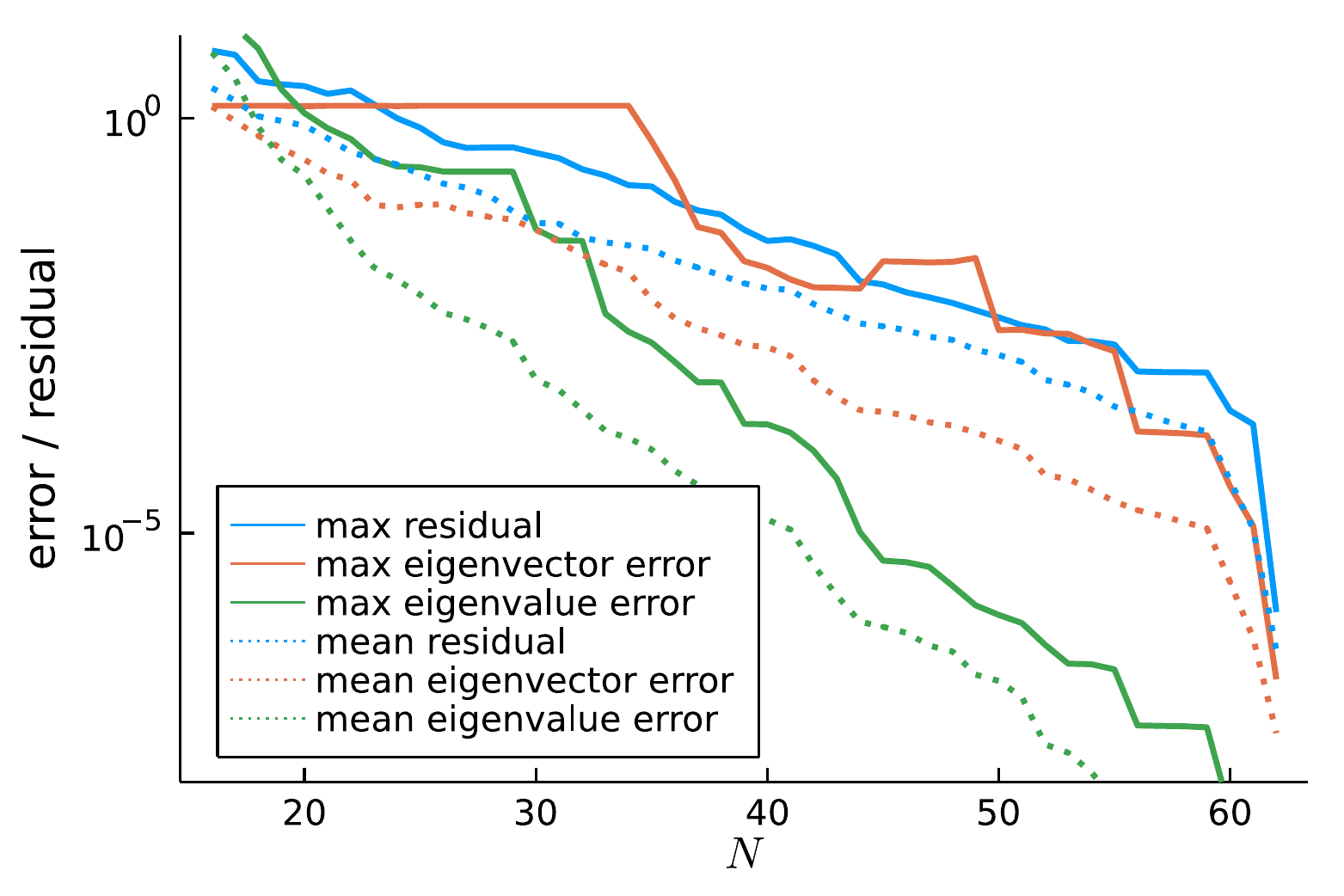}
    \caption{
    Decay of the error quantities of the RBM and the residual during the greedy
    algorithm with respect to $N$ for the triangular lattice model with
    $\nx=2$, $ \ny=2$.
    A uniformly spaced grid of $20\times 20\times 20$ points was used
    for training and a testing grid of $19\times 19\times 19$ points
    distinct to $\Xitrain$.
    The mean value of the corresponding quantity over the test-grid $\Xitest$
    are dotted.
    }
    \label{fig:Triangle_error}
\end{figure}

% We now shed the attention to a finer test-grid $\Xitrain$ with 
For an RBM trained on the $\Xitrain$
the decay of the approximation errors for different values of $N$
is reported in Figure~\ref{fig:Triangle_error} using a test-grid with
$19\times 19\times 19$ points different to $\Xitrain$.
We can clearly observe a stagnation of the eigenvector error $\vecerror$,
measured as in~\eqref{eq:vecerror} in the early phase of the greedy algorithm.
This can be explained as follows. For small numbers $N$ of basis functions, the
approximability of RBM is still quite inaccurate and might even fail to get the
degeneracy right. As a result, the error in the density matrix $\PhirbN(\bm\mu)
\PhirbN(\bm\mu)^\dagger$ is of order one. As the number $N$ increases, this
issue is cured and we observe the normal behavior of error reduction.
In order to shed more light on this point, we also report in
Figure~\ref{fig:Triangle_error} the mean values of the error quantities over
the test-grid $\Xitest$, i.e. replacing  $\max_{\bmu\in\Xitest}$ by an average
over $\Xitest$, and we observe that the stagnation of the error does not occur
in this mean quantities, which implies that the wrong prediction of the
degeneracy only occurs within a small subset of points in $\Xitest$.

% Finally, Figure~\ref{fig:Triangle_occ_map} shows the maximal occupation number, defined as 
% \[
%      \frac{1}{m}\max_{n=1,\ldots,{\mathcal N}}  \Big(\PhirbN(\bm\mu) \PhirbN(\bm\mu)^\dagger\Big)_{n,n}
% \]
% for degenerate states for varying $\mu_1=J_1/J_3$ and $\mu_2=J_2/J_3$ and fixed $J'=0.01$ and $J'=0.1$ respectively. 

% \BS{Todo: Speak about degeneracies here: where they are}

% \SW{where $m$ denotes the degeneracy (???) There was before an $m$ defined as $2^{\nx\ny}$. If this is meant here, we should define it here. Not clear to me, how to interpret the plot. Do we  need it? Is it mainly saying that the occupations are  overall rather low and further reduced by the degeneracy (I guess). This indicates that the ground states are not close to any of the basis states, is this the point to make here? }
% \BS{add something about degeneracy}

%%%%%% Triangle: error decay %%%%%% 
%%%%%% end figure %%%%%% 

%%%%%% Triangle: error map %%%%%% 
\begin{figure}[t!]
    \centering
    \begin{overpic}[width=0.45\textwidth]{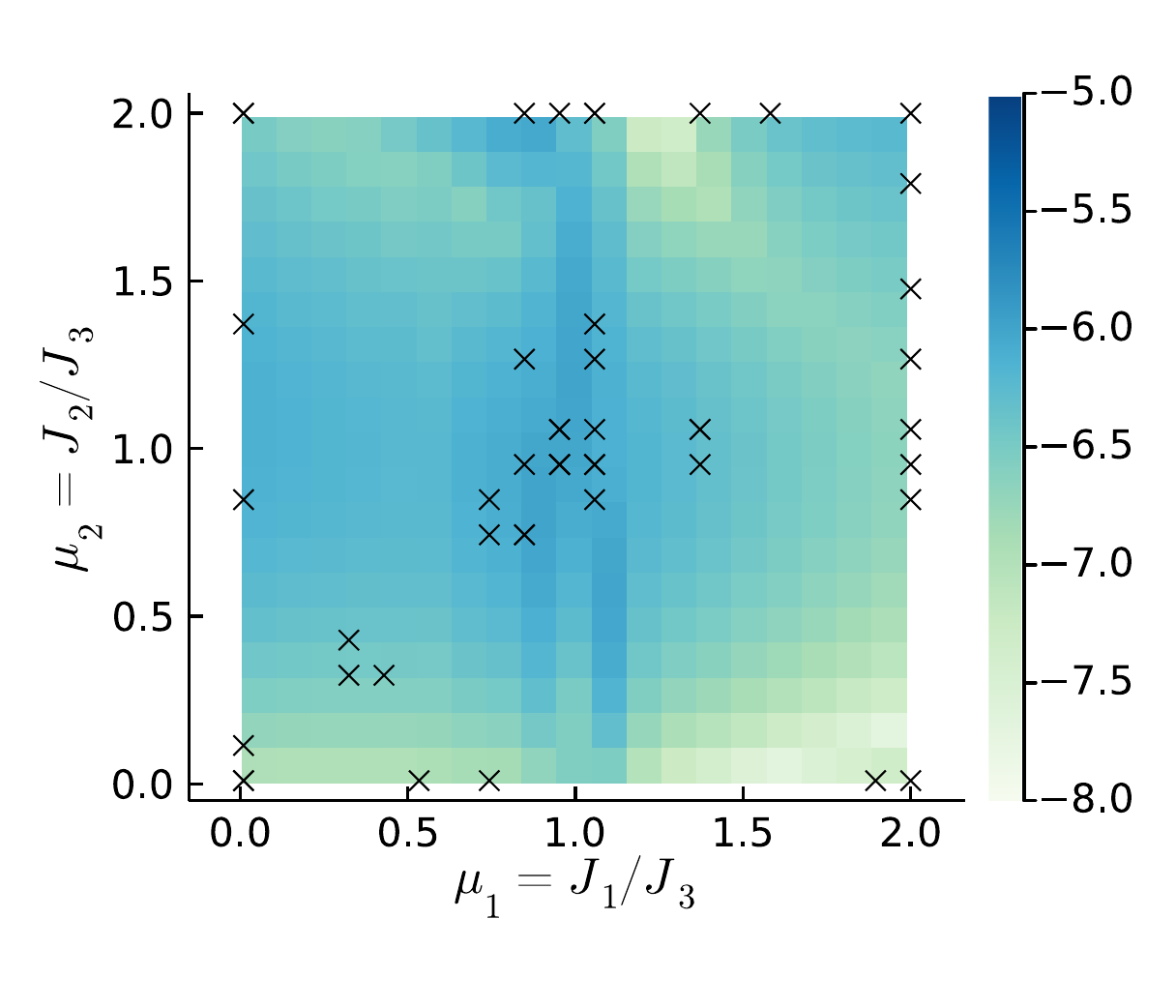}
    \put (87,12) {$\log_{10}$}
    \end{overpic}
    \caption{
    Logarithmic distribution of the eigenvalue-error of the RBM for $N=50$ over
    the parameter domain and the sample points selected by the greedy (crosses)
    for the triangle model with $\nx=2$, $\ny=2$.
    The maximal error for across all values of $\mu_3 = J' / J_3$
    is shown and sampling points are projected onto the $\mu_1$-$\mu_2$ plane.
}
    \label{fig:Triangle_error_map}
\end{figure}
%%%%%% end figure %%%%%% 

\begin{figure}[t!]
    \centering
    \begin{overpic}[width=0.45\textwidth]{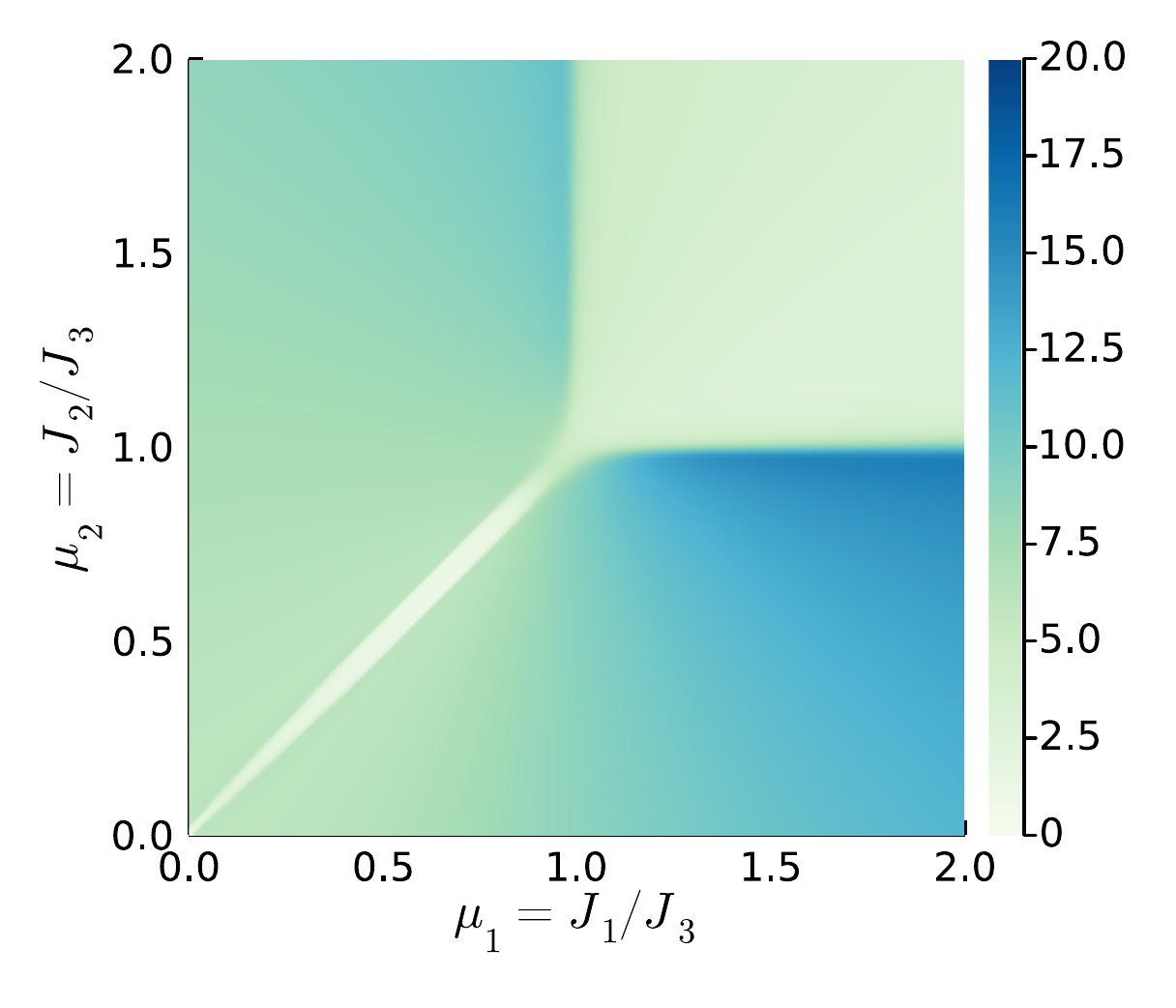}
    \put (0,75) {(a)}
    \put (45,83) {$(0,\pi)$}
    \end{overpic}
    \begin{overpic}[width=0.45\textwidth]{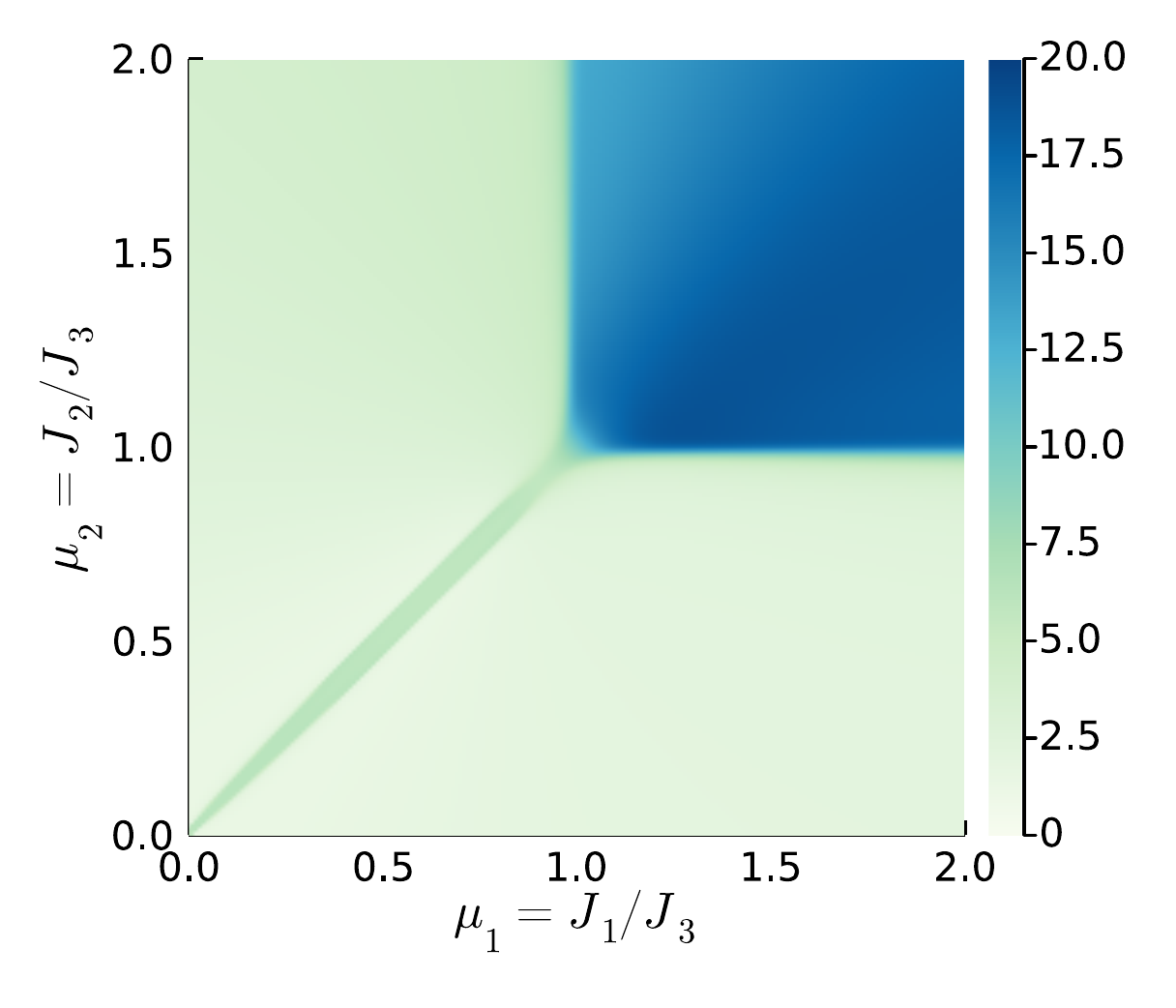}
    \put (0,75) {(b)}
    \put (45,83) {$(\pi,0)$}
    \end{overpic}
    \caption{
        The structure factor of the triangle model with $\nx=2$, $\ny=2$ and
        $J'/J_3=0.1$ at the k-points $(0,\pi)$ (a) and $(\pi,0)$ (b) for $N=62$.
        }
    \label{fig:Triangle_sf_map}
\end{figure}

Physically, what happens in this quantum spin model throughout most of the
considered parameter space is that each triangle forms a single effective
spin-$\frac12$ degree of freedom (as long as $J_1$, $J_2$ and $J_3$ are not all
equal), and these moments are then coupled by the inter-triangle exchange terms
$J'>0$ to form an ordered ground state in the thermodynamic
limit~\cite{WesselHaas2001,Wang2001}. The onset of the corresponding spin
ordering pattern can already be identified by examining the structure factor
$\calS(\bmu;\bk)$ on comparably small system sizes. Of particular interest are
the values of $\calS$ at the specific momenta $\bk=(0,\pi)$
and $(\pi,0)$, since these mixed ferromagnetic-antiferromagnetic states occupy
most of the considered parameter regime $\bmu$, and we concentrate here on the
results for $\nx = \ny =2$, for which both these momenta are present for
the considered periodic boundary conditions.
For this model our greedy approach allows to construct small, but accurate
reduced basis models. For example with only $N = 50$ basis vectors an eigenvalue
error below $10^{-6}$ is obtained across the full parameter range,
compare Figure~\ref{fig:Triangle_error_map}.
Based on an RBM with $N=62$ we thus consider in Figure~\ref{fig:Triangle_sf_map}
the value of the structure factor at the $\bk=(0,\pi)$ and $(\pi,0)$ momenta
for a value of $J'/J_3=0.1$. We observe in particular the predominance of the $(\pi,0)$
ordering in the upper right part of the considered parameter regime and along
the line $\mu_1=\mu_2$. An inspection of the full momentum-resolved structure
factor (not shown here) indeed verifies that no other momenta provide a more
dominant contribution to the magnetic structure factor. For the considered
system size, the remainder of the $\mu_1$-$\mu_2$-plane is dominated by the
rotated $(0,\pi)$ state. We note that there is only a rather narrow region
around the isotropic point $J_1=J_2=J_3$, for which a fully antiferromagnetic
state $(\pi,\pi)$ prevails on larger lattices~\cite{WesselHaas2001,Wang2001}.

%%%%%% Triangle: SF map %%%%%% 
%%%%%% end figure %%%%%% 

We also tested our greedy RBM approach on larger system sizes of the triangular lattice
model, namely $\nx = 2$, $\ny = 3$ as well as $\nx = 3$, $\ny = 2$. Both settings feature
a Hilbert space dimension beyond 250\,000. Nevertheless an RBM model with $N = 163$
--- less than a thousandth of the Hilbert space size --- is sufficient to obtain an
eigenvector residual below $0.1$ and a qualitative representation of the structure
factor. % throughout a test grid of $19\times 19\times 19$ points.
However, note that
% We skip a more detailed presentation of computational results, since
these system sizes are less suitable to examine
which ordered phase emerges in the thermodynamic limit: systems with an odd
number of unit cells in periodic boundary conditions give rise to frustration of the
antiferromagnetic alignment of the trimer spins along the directions with an odd
number of repeats.

Finally, let us illustrate to what extent the reduced basis is useful as an
initial guess for the computation of truth solutions. Indeed, it allows us to
save noticeably on the number of iterations required to obtain the converged
ground states. For the triangle model with $\nx = \ny = 2$, for example,
a naive guess (using a previously computed neighboring point on the grid $\Xitest$)
requires around 18\,000 LOBPCG iterations to converge all ground states
on our test grid of size $9^3=729$.
In contrast employing the reduced basis with $N = 62$
as a guess one obtains all ground states
using only a total number of around 1\,600 iterations,
again for all points in the grid.
It turns out that the accuracy of the reduced basis is important though.
For example, if only $N=50$ is taken only 25\% iterations are saved,
i.e.~around 13\,000 iterations are needed.

%%%% Conclusions %%%%%%%%
\section{Conclusions}
\label{sec:Conclusion}

In this article, we applied the reduced basis method (RBM) to parametrized quantum spin systems.
Such systems naturally lead to Hamiltonians which feature an affine decomposition.
In this work we exploit this structure in the context of the RBM in order to explore
phase diagrams and other output quantities in a complexity that is independent of
the dimension of the Hilbert space.
The required reduced basis is built up using a greedy strategy based on the residual
as an error surrogate, requiring only a small number of exact ground state computations
(around 20) to already reach a qualitative results.
We tested the methodology for two systems, a one-dimensional chain of excited
Rydberg atoms and a geometrically frustrated antiferromagnetic two-dimensional
spin-$\frac12$ lattice model.

This proof-of-concept was already quite conclusive.  First, at the range of
number of sites considered in this article, the manifold of ground-states can
be approximated by a surprisingly low-dimensional approximation space.  From a
theoretical viewpoint it would be interesting to understand the growth of the
effective dimension of the solution manifold for larger numbers of particles.
From a numerical perspective, the greedy strategy which uses the residual of
the eigenvalue problem as error surrogate turned out to be reliable in
generating reduced basis spaces in the considered tests cases.  Further, the
greedy algorithm was able to deal with degenerate ground-states while being
very efficient in the number of high-dimensional computations. Accurate reduced
bases were assembled and the decay rate of the error is similar to the
(optimal) one from the singular values.  The reduced basis method was able to
reproduce output functionals, such as structure factors or occupation numbers,
fairly accurately and, e.g., only 20 basis functions (thus requiring only 20
expensive truth computations) were needed to provide a qualitatively correct
occupation number plot in the case of the Rydberg chain model.  Further, the
method was surprisingly accurate in describing the sharp transition between the
$\mathbb Z_3$ and $\mathbb Z_4$ crystalline phase with so few basis functions.

To what extend this favorable reduction of computational cost
enabled by the greedy strategy generalizes to other quantum spin models
is an interesting direction for future work.
In particular whether one can expect the RBM to perform well on systems,
in which the degree of entanglement varies
across the parameter space, is an open research problem.
Our expectation is that this
should not be significant with respect to the effectiveness of the RBM 
provided that the small corner of the Hilbert space that is actually
entangled within the ground-state wave-function does not
vary drastically under small parameter variations.
In this case a small number of truth solves should be sufficient to capture
each relevant Hilbert space corners across the parameter domain.
Moreover we emphasize that even on unseen systems the greedy strategy
can be employed readily,
since a careful monitoring of the decay of the residual
gives direct insight whether the obtained reduced basis provides
a good approximation within the studied parameter space.

One of the current limitations of our implementation, but not of the methodology itself, seems to be the use of standard eigensolvers such as the LOBPCG-method to compute the (truth) ground-states. 
As a perspective, we aim in future work to perform this task via more sophisticated tensor network methods, which efficiently deal with a larger number of microscopic constituents.
They can in principle be embedded in a natural way as a black-box solver within the RBM framework, since they easily give access to the scalar products and matrix elements needed to construct the RBM surrogate model.

Another potential application of this methodology --- besides the fast scan of
phase-diagrams --- is the use of the surrogate model in order to generate
initial guesses for more advanced eigenvalue solvers at a generic point in
parameter space.
We actually already exploited this strategy in our implementation, whenever we
needed to compute a truth solution and a reduced basis was available.
We demonstrated an initial guess provided by an accurate reduced basis to lead to
10 times fewer iterations in the eigenvalue solver compared to the naive approach
of employing the solution of a neighboring parameter value.

%%%% Acknowledgments %%%%%%%%
\section*{Acknowledgments}
The authors would like to thank the Center for Simulation and Data Science (CSD) of the J\"ulich Aachen Research Alliance which provided the opportunity to meet and discuss across the different disciplines. 
MR acknowledges partial support from the Deutsche Forschungsgemeinschaft (DFG), project grant 277101999, within the CRC network TR 183 (subproject B01), and the European Union (PASQuanS, Grant No. 817482). SW acknowledges
support by DFG through Grant No. WE/3649/4-2 of the FOR 1807 and through RTG 1995.

\bibliography{biblio}

\end{document}